\documentclass[twocolumn]{article}

\usepackage{geometry}
\usepackage{natbib}

\usepackage{amsfonts}
\usepackage{amsmath}
\usepackage{color}

\usepackage{algpseudocode} 
\usepackage{float}
\usepackage{import}
\usepackage{multicol} %

\usepackage{xcolor}
\usepackage{todonotes}
\usepackage{amssymb}

\definecolor{mon_rouge}{HTML}{AD4B42}
\definecolor{mon_bleu}{HTML}{4F6A9C}
\definecolor{mon_vert}{HTML}{82A35D}
\definecolor{mon_jaune}{HTML}{DCB355}

\usepackage{array}
\newcolumntype{C}[1]{>{\centering\let\newline\\\arraybackslash\hspace{0pt}}m{#1}}

\usepackage{xparse}
\usepackage{xspace}

\newcommand{\symb}[1]{{\ensuremath{#1}}\xspace}

\DeclareDocumentCommand \argmax { s D[]{} m } {
	\IfBooleanTF{#1}
		  {\symb{\operatorname*{arg\,max}_{#2} #3}}
		  {\symb{\operatorname{arg\,max}_{#2} #3}}
}

\DeclareDocumentCommand \argmin { s D[]{} m } {
	\IfBooleanTF{#1}
		  {\symb{\operatorname*{arg\,min}_{#2} #3}}
		  {\symb{\operatorname{arg\,min}_{#2} #3}}
}

\DeclareDocumentCommand \newsymbol {m m D[]{} D<>{} d()} {
	\DeclareDocumentCommand #1 { t^ s d[] d<> d() }{%
		\symb{ %
		\IfBooleanTF{##2}%
		{\IfBooleanTF{##1}{\hat{#2}}{#2}%
			_{\IfNoValueTF{##3}{}{##3}}%
			^{\IfNoValueTF{##4}{}{##4}}%
			\IfNoValueTF{##5}{}{(##5)}
		}%
		{\IfBooleanTF{##1}{\hat{#2}}{#2}%
			_{\IfNoValueTF{##3}{#3}{##3}}%
			^{\IfNoValueTF{##4}{#4}{##4}}%
			\IfNoValueTF{##5}%
				{\IfNoValueTF{#5}{}{(#5)}}%
				{(##5)}%
		}}%
	}%
}

\newsymbol{\p}{p}()

\DeclareDocumentCommand \set { m d[] }{%
	\symb{\{#1\}_{{\IfNoValueTF{#2}{}{#2}}}}%
}

\newcommand{\eql}[1]{\label{eq:#1}}

\newcommand{\fig}[1]{Figure~\ref{fig:#1}}

\newcommand{\secref}[1]{Section~\ref{sec:#1}}
\newcommand{\secl}[1]{\label{sec:#1}}

\newcommand{\mr}[1]{\symb{\mathbb{R}^{#1}}}

\newcommand{\ie}{\textit{i.e.}\xspace}
\newcommand{\eg}{\textit{e.g.}\xspace}

\newcommand{\done}[1]{}
\newcommand{\elsewhere}[1]{}

\newsymbol{\O}{\mathcal{O}}
\newsymbol{\Th}{\Theta}
\newsymbol{\o}{o}

\newcommand{\matthijs}[1]{{\color{mon_rouge}\small #1}}

\usepackage{hyperref}

\makeatletter
\def\therule{\makebox[\algorithmicindent][l]{\hspace*{.5em}\vrule height .75\baselineskip depth .25\baselineskip}}%

\newtoks\therules%
\therules={}%
\def\appendto#1#2{\expandafter#1\expandafter{\the#1#2}}%
\def\gobblefirst#1{%
  #1\expandafter\expandafter\expandafter{\expandafter\@gobble\the#1}}%
\def\LState{\State\unskip\the\therules}%
\def\pushindent{\appendto\therules\therule}%
\def\popindent{\gobblefirst\therules}%
\def\printindent{\unskip\the\therules}%
\def\printandpush{\printindent\pushindent}%
\def\popandprint{\popindent\printindent}%

\algdef{SE}[WHILE]{While}{EndWhile}[1]
  {\printandpush\algorithmicwhile\ #1\ \algorithmicdo}
  {\popandprint\algorithmicend\ \algorithmicwhile}%
\algdef{SE}[FOR]{For}{EndFor}[1]
  {\printandpush\algorithmicfor\ #1\ \algorithmicdo}
  {\popandprint\algorithmicend\ \algorithmicfor}%
\algdef{S}[FOR]{ForAll}[1]
  {\printindent\algorithmicforall\ #1\ \algorithmicdo}%
\algdef{SE}[LOOP]{Loop}{EndLoop}
  {\printandpush\algorithmicloop}
  {\popandprint\algorithmicend\ \algorithmicloop}%
\algdef{SE}[REPEAT]{Repeat}{Until}
  {\printandpush\algorithmicrepeat}[1]
  {\popandprint\algorithmicuntil\ #1}%
\algdef{SE}[IF]{If}{EndIf}[1]
  {\printandpush\algorithmicif\ #1\ \algorithmicthen}
  {\popandprint\algorithmicend\ \algorithmicif}%
\algdef{C}[IF]{IF}{ElsIf}[1]
  {\popandprint\pushindent\algorithmicelse\ \algorithmicif\ #1\ \algorithmicthen}%
\algdef{Ce}[ELSE]{IF}{Else}{EndIf}
  {\popandprint\pushindent\algorithmicelse}%
\algdef{SE}[PROCEDURE]{Procedure}{EndProcedure}[2]
   {\printandpush\algorithmicprocedure\ \textproc{#1}\ifthenelse{\equal{#2}{}}{}{(#2)}}%
   {\popandprint\algorithmicend\ \algorithmicprocedure}%
\algdef{SE}[FUNCTION]{Function}{EndFunction}[2]
   {\printandpush\algorithmicfunction\ \textproc{#1}\ifthenelse{\equal{#2}{}}{}{(#2)}}%
   {\popandprint\algorithmicend\ \algorithmicfunction}%
\makeatother

\newif\ifRR
\RRfalse

\begin{document}

\makeatletter
\g@addto@macro\@maketitle{
  \begin{figure}[H]
  \setlength{\linewidth}{\textwidth}
  \setlength{\hsize}{\textwidth}
  \centering
\hspace*{-1.5cm}
\includegraphics[width=1.15\linewidth]{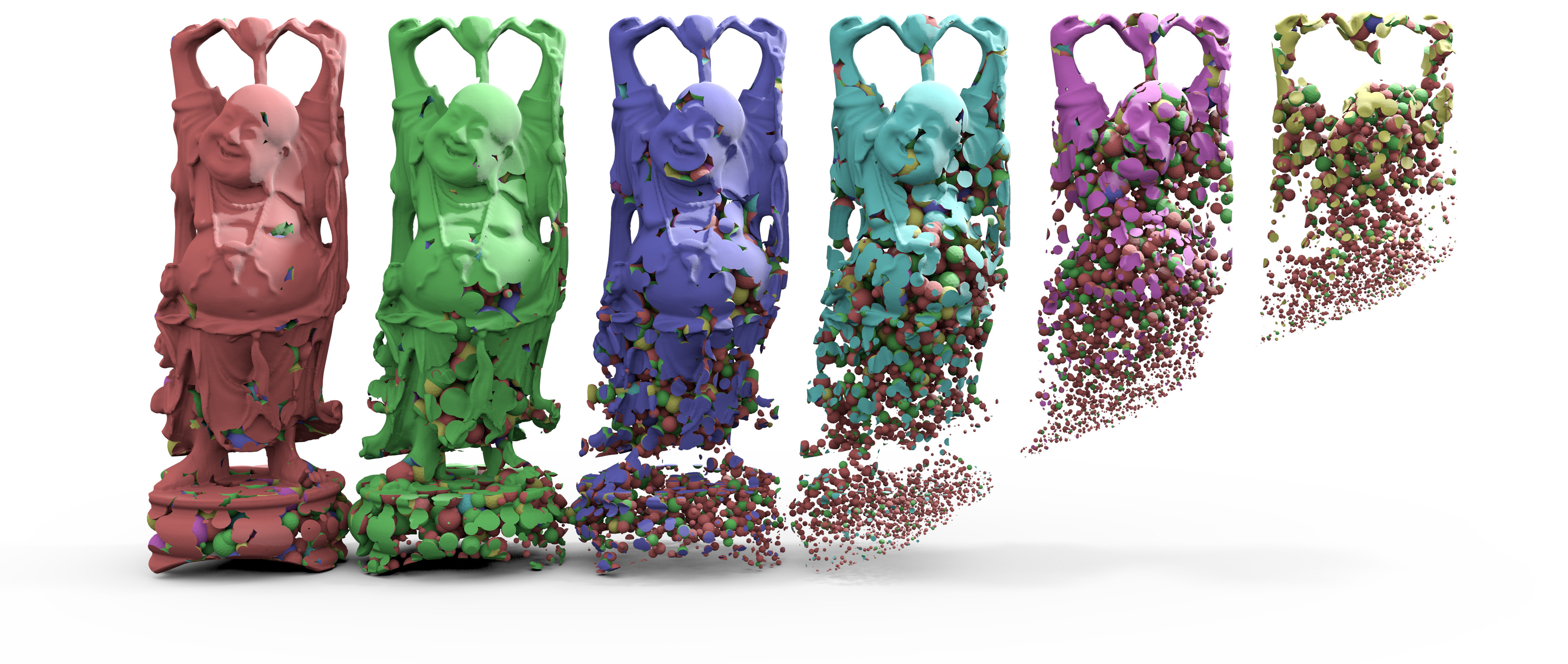}
\vspace{-1.2cm}
\caption{\label{fig:6buddha}
Intersection of 6 Buddhas with the union of 100,000 spheres (total 24 million triangles). Computed in 8~seconds on a desktop machine.
}
  \end{figure}
}
\makeatother

\markboth{M. Douze et al.}{}

\title{
	\makebox{QuickCSG: Fast Arbitrary Boolean Combinations of N Solids}
}

\author{Matthijs Douze, Jean-S\'ebastien Franco, Bruno Raffin}

\maketitle

\begin{abstract} 
QuickCSG computes the result for general N-polyhedron boolean expressions without an intermediate tree of solids. We propose a vertex-centric view of the problem, which simplifies the identification of final geometric contributions, and facilitates its spatial decomposition. The problem is then cast in a single KD-tree exploration, geared toward the result by early pruning of any region of space not contributing to the final surface. We assume strong regularity properties on the input meshes and that they are in general position. This simplifying assumption, in combination with our vertex-centric approach, improves the speed of the approach. Complemented with a task-stealing parallelization, the algorithm achieves breakthrough performance, one to two orders of magnitude speedups with respect to state-of-the-art CPU algorithms, on boolean operations over two to dozens of polyhedra. The algorithm also outperforms GPU implementations with approximate discretizations, while producing an output without redundant facets. Despite the restrictive assumptions on the input, we show the usefulness of QuickCSG for applications with large CSG problems and strong temporal constraints, \eg modeling for 3D printers, reconstruction from visual hulls and collision detection.
\end{abstract}

\section{Introduction}

Solid modeling using boolean operations is an emblematic problem in computer graphics and computational geometry, almost as old as these research topics themselves. %
It has found its way in every solid modeler in the industry, whether applied to model design for aviation, transportation, manufacturing, architecture, or entertainment. It is also an ubiquitous building block and subject of interest for many fields of research, including computer graphics, computer vision, robotics, virtual reality, and generally any topic where geometric models of subjects of interest are to be manipulated, constructed, truncated or combined.

Since the first introduction of boundary representations (B-Rep) \cite{baumgart74}, the problem has received considerable attention and been the subject of extensive work over more than 40 years. It is all the more striking that, despite the many existing algorithms and variants in this huge corpus, the vast majority of algorithms rely on a common principle and canvas found in the earliest formalizations of the problem~\cite{requicha85,laidlaw86},  summarized hereafter. First and foremost, boolean B-Rep merging algorithms are written for the case of two solids. Second, the computation is divided in three stages: an initial \textit{subdivision} stage, where the boundaries of both objects are split in two component groups along their intersection with the other object's boundary. A \textit{classification} stage follows, where each group is classified as belonging inside or outside the other object. In the final \textit{reconstruction} stage, the relevant primitives are gathered and connected to build the final model in accordance to the boolean expression. Note that the subdivision and classification require to intersect and situate all primitives of an object's boundary with respect to the primitives of the other object's boundary, which if done naively leads to impractical quadratic-time algorithms. Thus, a third  aspect of most algorithms is the use of spatial decomposition structures, often hierarchical, to enable sublinear $\mathcal{O}(\log m)$ access to each of the $m$ object primitives. The construction of this data structure  becomes the bottleneck of the algorithm, giving it its typically quasilinear time complexity $\mathcal{O}(m \log m)$ in the number of input primitives~\cite{naylor90,hachenberger07}. An inherent drawback of this dominant approach is that all input solid's primitives are fully decomposed, but typically only a fraction of those primitives contribute to the output result; the time spent computing hierarchical decompositions for non-contributing primitives is thus useless and can be eliminated, as we will show.

\begin{figure*}
\includegraphics[width=\linewidth]{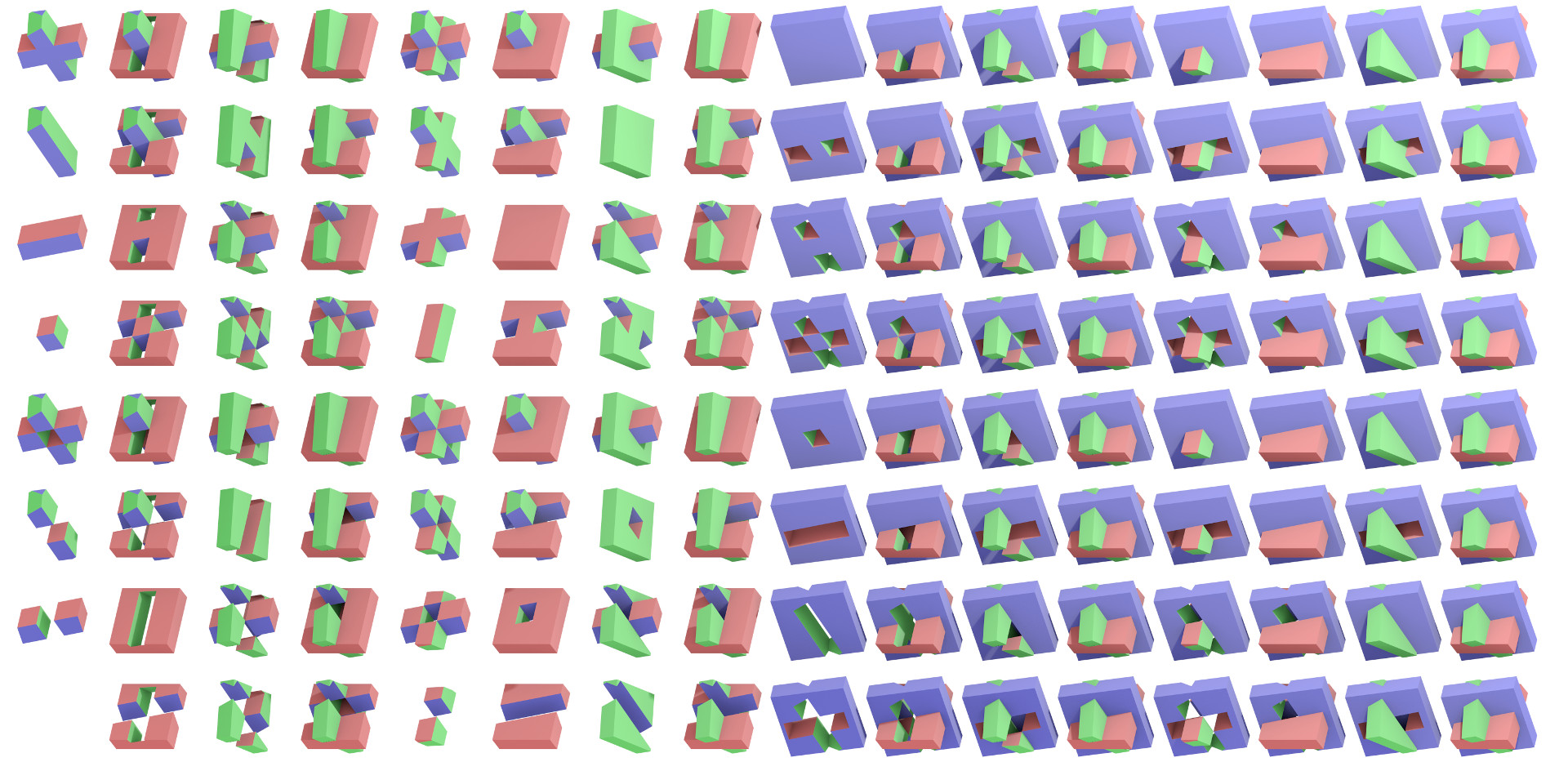}
\caption{\label{fig:boxtable}
   There are $2^{2^3}=256$ possible boolean functions of three inputs, of which we show 128, applied to 3D cuboids (the $128$ other ones are the same with inside and outside flipped). 
   In our figures, each input solid is assigned a color, which is inherited by the output facets it contributed to the result.
   Some functions cannot be computed from binary CSG operations without producing degeneracies. For example, the 2nd solid in reading order represents the masking operation $(\mathcal{P}_1 \cap \mathcal{P}_2) \cup (\mathcal{P}_3 \backslash \mathcal{P}_2)$ with $\mathcal{P}_2$ in green,  $\mathcal{P}_3$ in red and $\mathcal{P}_1$ in blue. The union operation in this expression is degenerate because facets of $\mathcal{P}_2$ appear on both sides. 
}
\end{figure*}

The more general case of $n$ input solids is classically addressed by combining pairwise operations in arbitrary boolean expressions. The approach has been formalized as Constructive Solid Geometry~(CSG)~\cite{requicha80,mantyla87}, where these expressions map to CSG-trees of boolean operations. Evaluating the result for B-rep solids then relies on combining two boundaries at each node of the tree using an existing two-solid boolean algorithm. This approach however has some significant drawbacks, as they involve computing and storing a set of intermediate results, which are then recombined until the tree expression is fully resolved. The successive recombinations can be error prone, in fact leading to inherently degenerate situations for some expressions. Figure~\ref{fig:boxtable} shows for example all possible boolean function outcomes for the case of three solids and illustrates such cases. Another inherent limitation is that certain expressions lead to combinatorial size trees which are impractial to compute. Consider for example identifying the solid whose volume is the intersection of at least $k$ solids, an operation later referred as $min\!\!\!-\!\!\!k$: this operation involves computing the union of all possible intersections over $k$ or more solids, yielding a combinatorial CSG tree size.

In this paper, we challenge these dominant views of boolean modeling to eliminate these drawbacks. First, our algorithm directly computes the result of arbitrary $n$-ary boolean expressions, avoiding the overhead and intermediate results of binary CSG-tree approaches.
This is achieved by directly identifying vertices of the final polyhedron among all relevant vertex candidates, with an efficient vertex classification test using bitvector evaluations of the $n$-ary CSG function.

Second, our algorithm performs the classification and subdivision stages simultaneously: as opposed to existing approaches, our hierarchical decomposition and exploration is performed on the combined set of all input solids primitives and guided by their participation to the final solid B-rep. 

Both key aspects of our algorithm are made possible by embedding all input solids in a single KD-tree exploration. The algorithm is able to retrieve the final result directly because each new KD-tree node explored is classified as soon as it is created, by inferring whether its contents is completely inside or outside the final solid, or if it may participate to the final solid boundary instead. %
The KD-tree is only subdivided in the latter case, pruning large sets of primitives that do not need additional work, and focusing all computational effort and refinement on those space cells containing intersecting primitives participating to the final result.

Consequently, the depth and size of our KD-tree depends on the output model size instead of input model size, leading to gains in time and space complexity that grow with the number of input solids and the input-to-output primitive ratio. This means that our algorithm also outperforms the traditional two-solid boolean algorithms as soon as the result size is smaller than the input size. Finally, as our KD-tree decomposes space into non-intersecting cells that can be processed independently, the classification stage and subsequent subdivision and reconstruction stages naturally lend themselves to parallel evaluation, further substantiating the temporal gain over all state-of-the-art polyhedral boolean evaluation algorithms tested, including recent GPU implementations. 

~\\
\textbf{Contribution summary.}
\begin{itemize}
\item Complexity: we improve asymptotic upper bounds for running time and memory consumption, both analytically and experimentally, with  
a simple and parsimonious output-sensitive algorithm;

\item Any boolean operation: 
QuickCSG directly computes the result of arbitrary boolean expressions over a set of input solids, certain classes of which cannot be realistically computed with existing binary boolean approaches;

\item Applications on practical use cases: solid modeling for 3D printing, collision detection, and silhouette-based 3D reconstruction from visual inputs;

\item Benchmark: we introduce a benchmark comprised of several dozen datasets, made available to the research community with our implementation.

\end{itemize}

The paper is organized as follows. We first review state of the art of boolean operations (\secref{related}), then propose a formalization of the N-solid boolean problem (\secref{geometry}). We then explain how vertices (\secref{finalverts}) and facets (\secref{finalfacets}) of the final polyhedron can be directly identified. We show how the problem can be cast as a KD-tree divide-and-conquer exploration (\secref{kdtree}), and how the algorithm and this exploration in particular can be performed in parallel (\secref{parallel}). We validate the algorithm experimentally in \secref{results} and discuss applications in \secref{applications}.

\section{Related Work}
\secl{related}

\subsection{Boolean Solid Modelling Background}

In the 1970's, boolean solid modeling and boundary representations (B-Reps) have been simultaneously pioneered in the context of computer graphics~\cite{braid75} and computer vision~\cite{baumgart74}. Both proposed discrete representations of solid boundaries as a conjunction of simpler polygonal primitives, either winged edges (Baumgart) or loops (Braid).
While many of the ideas are already present in Braid's work, the idea that solids could be specified as a tree of boolean operations (Constructive Solid Geometry or CSG) was theorized by~\cite{requicha80}, and various practical implementations proposed for polyhedral boundaries~\cite{requicha85,laidlaw86}, in particular setting the standard for the aforementioned 3-stage canvas. Robustness was by then identified as a recurring issue, due to lack of formal description of degenerate solid configurations, and numerical computation error in near-coincident situations. Most algorithms thereafter, including industrial implementations, thus conform to a set-based formalism with algebraically closed regularized boolean set operations~\cite{requicha77}, ensuring results exclude any non-volume enclosing (dangling) surface primitives. Several works also took on the task of painstakingly accounting for all degenerate relative configurations of solid primitives~\cite{hoffmann89,mantyla87}, leading to tedious algorithm descriptions. They notably formalize the B-Rep primitive hierarchy as vertex, edges, faces and shells, and the two-polyhedra intersections and degeneracy cases as arising from the possible intersection combination of each primitive type of solid A to each primitive type of solid B. To avoid the complete enumeration, many implementations focus on generic triangle-to-triangle or polygon-to-polygon as their central intersection unit. Even with this simplification, the complexity of dealing with all cases is known to yield unreliable implementations, including in commercial software, as reported in various test cases~\cite{wang2011csg,feito13}. Our algorithm has a significantly simplified core that focuses all classification and subdivision efforts on producing the final output vertices, excluding higher order primitives or intermediate vertices. The final reconstruction stage  operates on these vertices, identifying the topologically correct final edges and loops through a posteriori logical vertex-to-vertex reconnections. This improves both the clarity and regularity of the proposed algorithm, and paves the way for tackling the otherwise unaffordable exact topology retrieval in the general N-polyhedron case.

Robustness has remained a dominant issue, with various solutions proposed reviewed by e.g.~\cite{hoffman01,yap04}, %
such as geometric predicate analysis %
 and fixed or arbitrary precision exact arithmetics. %
This effort has culminated with Hachenberger's work on CGAL~\cite{hachenberger07}, which uses arbitrary precision arithmetic, %
with the particularity that it follows Nef's formalism~\cite{nef78} instead of regularized booleans~\cite{requicha77}, i.e. it explicitly represents dangling primitives. While now standing out as a reference implementation of the research community, it is notoriously slow, and as most exact schemes, tedious to re-implement, leading to a somewhat paradoxical status: while the perception of the community is that the polyhedral B-Rep boolean problem is solved, free and commercial code is still being crafted and distributed using fragile but fast and memory-efficient geometric predicate evaluations. 

Most new contributions in this area are focusing on speeding up or easing the implementation of various algorithmic work cases of the classic two-polyhedron boolean algorithms, with e.g. new specialized data structures~\cite{kobbelt10}, faster exact arithmetic types~\cite{bernstein09}, optimization for particular inputs such as triangular meshes~\cite{feito13} or polyhedral cones of arbitrary basis~\cite{FrancoEPVH}. Each implementation relies on specific and non-optimal tradeoffs between implementation complexity, speed, memory footprint, input genericity, robustness (or lack thereof). We simultaneously improve over all problematic aspects: our greedy pruning scheme eliminates the need to compute any intermediate geometry and thus improves the complexity, robustness, memory footprint and execution time while enabling multi-arity boolean operations on $n$ polyhedra, including but not limited to binary boolean trees.  As this result strongly relies on the careful use of hierarchical subdivision structures, we  specifically review this aspect of prior art in the following section. %

\subsection{Subdivision Structures for Efficient Computation}

Because of the need for efficient subdivision and classification stages in the algorithm, a substantial research effort has been devoted to hierarchical structures in the context of boolean solid modelling. Some of the earliest axis-aligned plane-separation structures in this context are the polytrees~\cite{carlbom87} and extended octrees, which embed the polyhedral B-Rep primitives in their nodes~\cite{brunet90}. %
Binary space partitions (BSP) of polyhedral B-Reps were devised as a way to more efficiently store the polyhedron, where separating planes are based on input facets~\cite{thibault87}. The most common strategy to compute boolean combinations of two solids with these representations is to perform simultaneous traversal of both hierarchies to isolate intersecting primitives~\cite{brunet90}, or similarly compute a merged BSP tree itself representing the result~\cite{naylor90}.
Recent reference implementations continue to use variants of these seminal approaches, e.g. CGAL uses KD-trees as accelerated axis-aligned plane separating search structures~\cite{hachenberger07}, GTS uses axis-aligned bounding box (AABB) trees~\cite{popinet05}, while Carve CSG uses octrees~\cite{sargeant11}. A number of hybrid variants exist, which seek simultaneous benefit from the access simplicity of the octree structure and the representational flexibility of BSPs~\cite{adams03}.%

Of significant interest among such hybrid methods, \cite{pavic2010hybrid}  examine the boolean CSG binary tree with a single octree to embed all input geometry, subdivide cells down to a fixed cell size as long as two input solids are volumetrically present, then classify each leaf cell after subdivision by evaluating the CSG boolean tree expression. A key difference with our proposal is that the resulting meshes are stitched with an approximate local triangulation at intersecting leaf cells, while we compute true surface-to-surface boolean contributions, for arbitrary boolean expressions that need not be expressed with a boolean tree. \cite{feito13} uses a similar octree subdivision triggered by general two-surface presence, but focuses  only on triangular meshes and the two-solid case. Fundamentally for both approaches it can be noted that classification is still independently computed and not used to guide the subdivision.

A compelling case we make in this paper is that separating subdivision and classification stages, as done by all boolean B-Rep algorithms we are aware of, leads to an inherently suboptimal boolean algorithm. In light of the review of prior art, this is because the hierarchical structures proposed are in the vast majority of cases constructed as alternate representations of \textit{each individual input solid}, decomposing its geometric details with trees of logarithmic depth in that input solid's size. In contrast, our algorithm builds a single geometric decomposition, splitting nodes according to a partial classification of their content computed on the fly.  Branches not contributing  to  the resulting solid are pruned away, building a tree whose nodes are focused on the final surface geometry, with a depth logarithmic in the number of intersections present in the \emph{resulting solid} instead.
This yields two fundamental improvements over state of the art. First the algorithmic complexity is improved as it is now proportional to the logarithm of the resulting solid size. Second, because this tree classifies final contributions on the fly during subdivision, our algorithm stores only the information of the currently explored tree branch. This frees the algorithm from storing the  subdivisions structures of each input solid in preparation for a separate classification stage, as with previous methods.

\section{N-Polyhedron CSG Formalization}
\secl{geometry}
\label{sec:algo}

\newsymbol{\i}{i} %
\newsymbol{\j}{j} %
\newsymbol{\k}{k} %

\newsymbol{\ni}{0} %
\newsymbol{\nj}{0} %
\newsymbol{\nk}{0} %
\newsymbol{\yi}{1} %
\newsymbol{\yj}{1} %
\newsymbol{\yk}{1} %

\newsymbol{\x}{x} %
\newsymbol{\N}{n} %
\newsymbol{\P}{\mathcal{P}}[\i] %
\newsymbol{\F}{\mathcal{F}}[\i] %
\newsymbol{\V}{\mathcal{V}}[\i] %
\newsymbol{\I}{\mathbb{I}}[\i](x) %

\newsymbol{\f}{f}       %
\newsymbol{\Ires}{\mathbb{I}}[\f](x) %
\newsymbol{\Pres}{\mathcal{P}}[\f] %
\newsymbol{\Iv}{\mathbf{I}}(x)    %

\newsymbol{\b}{b} %
\newsymbol{\v}{v} %

This section introduces the representation of the input solids and the CSG operation.

\subsection{Definitions and Assumptions}

We consider  \N input polyhedra $\{\P\}_ {\i \in \{1, \cdots, \N\}}$, whose surfaces are assumed to be closed orientable 2-manifolds embedded in \mr{3}, \emph{i.e.} surfaces with no holes and with a consistent normal orientation. These classical assumptions ensure every polyhedron non-ambiguously defines a closed volume of \mr{3}. Each input polyhedron $\P = (\V, \F)$ is defined by its set of vertices \V and facets \F, each described as a loop of vertex indices whose order is consistent, \emph{e.g.}  given with counterclockwise orientation as seen from its outer region. We assume unicity of vertices, \emph{i.e.} no vertex coordinates are duplicated and adjacent loops share common vertices. A polyhedron may have various connected components. Facets are assumed convex and described with a single loop to simplify the explanation and implementation, although the reasoning  extends to general, non-convex polygons that may contain holes. 

The resulting shape may be complex, as any output facet may be shaped by arbitrary primitives of all inputs. The complexity of possible degeneracies between all types of primitives for two-polyhedron booleans is already quite daunting and error-prone to implement~\cite{hoffmann89,flaquer87}. Generalizing Hoffmann's analysis to \N-case degeneracies is not desirable, because the combinatorial possibilities of coincidental positioning of vertices, edges and facets are arbitrarily large. As an example, degenerate output vertices may arise from the coincidental positioning of anywhere from $4$ to \N input facets chosen among any input solid's facets, all of which would result in different special cases for reconstructing the vertex neighborhood. %
In practice, implementations most often get away with using double-precision floating-point arithmetic, as degeneracy cases are shown to be highly unlikely when dealing with noisy inputs, either resulting from an acquisition process~\cite{happybuddha96,FrancoEPVH} or artificially generated with jittering for this purpose. We follow this approach in QuickCSG.

\subsection{Boolean Functions of $\N$ inputs}

Instead of the usual CSG tree form of boolean expressions, we provide a framework for arbitrary expressions. We express a boolean solid operation using a boolean-valued function over \N boolean inputs, $\f: \{0,1\}^{\N} \rightarrow \{0,1\}$. We note $\I \in \{0, 1\}$ the indicator function of polyhedron \P, whose value reflects whether a point $\x \in \mr{3}$ is in polyhedron \P's inner volume. The indicator function \Ires of the final solid \Pres can then be computed using \f:
\begin{equation}
	\Ires = \f(\I[1], \cdots, \I[\N]).
	\label{eq:meshpos}
	\end{equation}
We define the \textit{indicator vector} of point \x as the tuple of its \N indicator functions, $\Iv = (\I[1], \cdots, \I[\N])$,  and  denote the CSG operation as occurring over its indicator vector, \textit{i.e.} $\Ires = \f(\Iv)$. Note that, as \P is given as a set of vertices and faces $(\V, \F)$, indicator function values \I can be computed %
by shooting a ray and counting the winding numbers of \x~\cite{geomtools4compgraph}. If a vertex is known to belong to the surface of a polyhedron, we denote the corresponding boolean value as `$s$', see Figure~\ref{fig:runningexample}. Note that we never compute function $f$ on $s$ inputs.

Any indicator function \Ires can be evalued from classical binary boolean operators (\eg as a conjunction of disjunctions), but alternative evaluations
are also possible  based on, for instance,  higher arity boolean operators or arithmetic operations. The rationale is to simplify the expression of some operations and  speed up the  evaluation. The following  examples show a few  operators whose n-ary formulation enables efficient evaluations:
\begin{alignat}{3}
	\textrm{\N-Intersection: }&& \I[\cap](x) &= \textrm{min}(\I[1],\cdots,\I[\N]),  \eql{union}\\
	\textrm{\N-Union: }&& \I[\cup](x) &= \textrm{max}(\I[1],\cdots,\I[\N]),  \eql{inter}\\
	\textrm{Mutual exclusion: }&& \I[\textrm{xor}](x) &= \I[1] \, \textrm{xor} \, \cdots \, \textrm{xor} \, \I[\N], \eql{xor} \\
	\textrm{In $k$ or more solids: }&&  \I[\textrm{min-}k](x) &= (\textstyle\sum_i \I) \ge k. \eql{mink}
\end{alignat}
 The min-$k$ operation retrieves the solid being part of at least $k$ input polyhedra. This operation in particular would be tedious to decompose over a binary CSG tree: it requires to evaluate the union of all possible intersections of $k$ solids, leading to a tree of combinatorial size.

Since the indicator vector can be efficiently represented as a bit vector stored in machine words, evaluating typical boolean functions \f is for most practical purposes a constant time operation in $\N$, either by directly evaluating a boolean test expression over the machine word, or by building lookup/hash tables for compatible expressions. 
Any binary tree of CSG operations can be expressed as a
single boolean function. Therefore, models built by binary CSG
operations can be directly constructed by our algorithm.
We will see below how these definitions are used to make final boundary surface decisions.

\section{Final Polyhedron Vertices}
\secl{finalverts}

\begin{figure}
\begin{center}
\def\svgwidth{5cm}
\raisebox{-0.5\height}{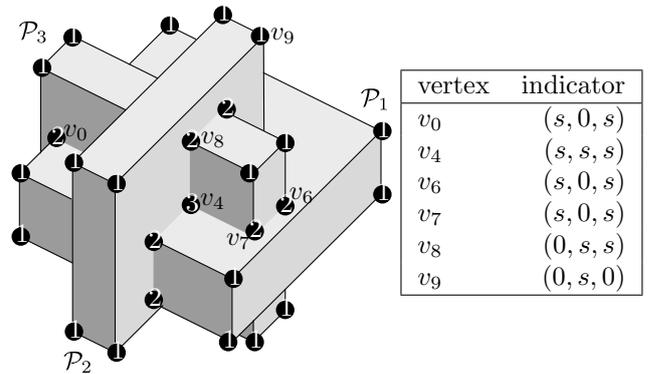}
\begin{tabular}{|lr|}
\hline
vertex & indicator \\
\hline
$v_0$ & $(s,0,s)$ \\
$v_4$ & $(s,s,s)$ \\
$v_6$ & $(s,0,s)$ \\
$v_7$ & $(s,0,s)$ \\
$v_8$ & $(0,s,s)$ \\
$v_9$ & $(0,s,0)$ \\
\hline
\end{tabular}
\end{center}
\caption{\label{fig:runningexample}
   Combination of three solids, with the orders of the vertices (in black circles). Left: the indicator vector for some of the vertices. 
}
\end{figure}

\begin{figure*}
\begin{center}
\makebox[0pt]{
\def\svgwidth{17.5cm}
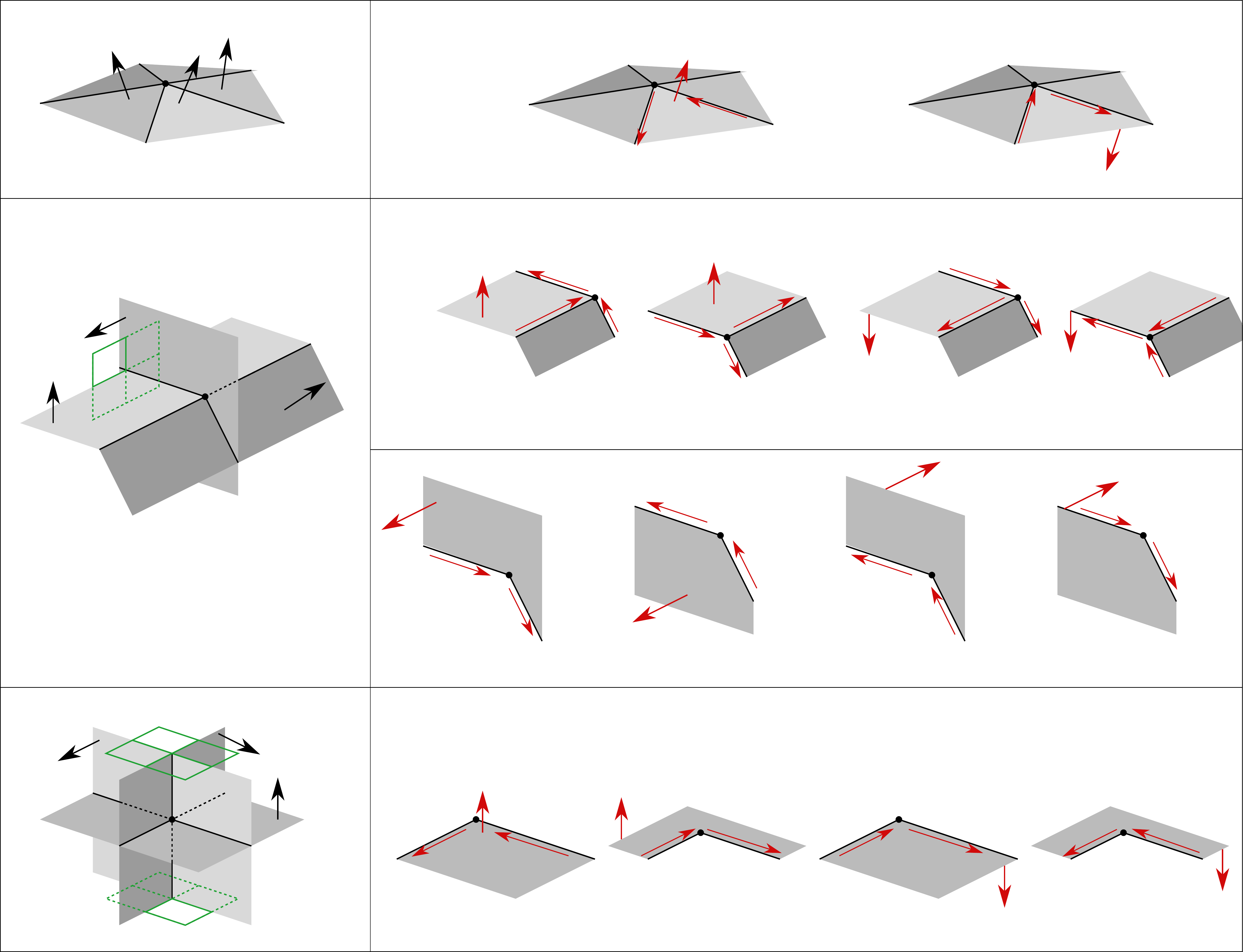
}
\end{center}
\caption{\label{fig:geom}
Vertex configurations and their corresponding possible looplets.
}
\end{figure*}

We now analyze the structure of the final polyhedron \Pres,  focusing on its vertices.
With primitives in generic (\emph{i.e.} non coincidental) position as assumed, vertices of the final polyhedron can be of only three types (Figure~\ref{fig:runningexample}): %
\begin{itemize}
\item \textit{First order vertices} are vertices already present in one of the input polyhedra \P's vertex set \V.
\item \textit{Second order vertices} result from the intersection of an edge of a polyhedron \P[i] and the facet of another polyhedron \P[j].
\item \textit{Third order vertices} result from the intersection of three facets of three different polyhedra \P[i], \P[j], and \P[k].
\end{itemize}
Thus, the order of a vertex is the number of $s$ bits in its indicator vector.
A trivial way to generate all possible vertex candidates is to examine all input vertices, edge-to-facet combinations, and three-facet combinations, and compute the resulting geometric intersections using standard algorithms~\cite{geomtools4compgraph}. Input primitives may intersect at various locations in space, without necessarily participating to the final surface, as determined by the CSG function. We thus propose a classification process to select the candidate vertices participating to the output result. Our description is illustrated on the left column of \fig{geom}, which summarizes the geometry of vertices of each order, and the notations used. In this figure, orientation information is given in red, and classification information in green. Small green grids are given to break down local subvolume configurations around the vertex $v$, and their corresponding classification bits, which are defined hereunder. 

\definecolor{darkgreen}{HTML}{1ea232}
\definecolor{darkred}{HTML}{d10a0a}

\newcommand{\Fb}[1]{\scalebox{0.8}{\color{darkgreen}{#1}}}
\newcommand{\Fbe}[2]{\Fb{$b_{#1} = #2$}}
\newcommand{\Fbt}[3]{\Fb{$t_{#1}^{#2} = #3$}}
\newcommand{\Fbb}[1]{\Fb{$b_{#1}$}}
\newcommand{\Fd}[1]{\color{darkred}{#1}}
\newcommand{\Fdd}[1]{\Fd{$d_{#1}$}}
\newcommand{\Fdf}[2]{\Fd{$F_{#1}^#2$}}
\newcommand{\FR}[1]{$\color{blue}{#1}$}

\subsection{Vertex Classification}

Intuitively a necessary condition for a vertex candidate to be kept is for it to lay at the border of the final solid, in other words it should be part of a surface transition from inside to outside \Pres. But this condition is not always sufficient.

The condition is sufficient for a \textbf{first order vertex.} By definition, the candidate vertex \v participates in the boundary of one input polyhedron \P, which separates the vicinity of the vertex into two subvolumes, inside and outside \P. For this vertex to lay on the boundary of \Pres, it must also partition the surrounding volume into inside and outside regions of \Pres. This information can be obtained by examining how \f(\Iv(\v)) transitions at the boundary of \P, \emph{i.e.} when the $i$-th bit of the indicator vector, initially a $s$ bit, is flipped between 0 and 1:
\begin{equation}
	\f(\I[1](\v), \cdots, 0, \cdots, \I[\N](\v)) \neq \f(\I[1](\v), \cdots, 1, \cdots, \I[\N](\v)). \eql{op1}
\end{equation}
This conditions is noted \texttt{isFinal1}$(\v)$, and tests wether there is a final indicator change when the boundary of \P is traversed. If the two expressions were equal, then the vertex \v would be completely inside (both 1) or outside (both 0) \Pres. This assumes all other bits \I[j](\v), with $j \neq i$, were computed for vertex \v by \emph{e.g.} ray shooting.

\newcommand{\iemph}[1]{\underline{#1}}
\newsymbol{\class}{\mathbf{b}}(v)

\textbf{Second order vertex.} Because facets of two polyhedra \P[i] and \P[j] are involved, the volume surrounding the vertex candidate is locally partitioned in four subvolumes, each of which may be decided to be inside or outside the final polyhedron \Pres by the CSG function~\f. We must therefore examine how each combination of boundary traversals at vertex $v$  influence the final indicator function, by evaluating the corresponding \i and $j$ bit-flippings of \Iv(v)~:
\begin{equation}
\begin{array}{rl}
	\b[\ni\nj]&=\f(\I[1](\v), \cdots, 0, \cdots, 0, \cdots, \I[\N](\v))\\
	\b[\ni \yj]&=\f(\I[1](\v), \cdots, 0, \cdots, 1, \cdots, \I[\N](\v))\\
	\b[\yi\nj]&=\f(\I[1](\v), \cdots, 1, \cdots, 0, \cdots, \I[\N](\v))\\
	\b[\yi\yj]&=\f(\I[1](\v), \cdots, 1, \cdots, 1, \cdots, \I[\N](\v)).
\end{array}
\end{equation}
We call $\class=(\b[\ni\nj], \b[\ni \yj], \b[\yi\nj], \b[\yi\yj]) \in \{0,1\}^4$ the classification vector of second order vertex \v.
Trivially, if all four bits turn out equal, the vertex \v is either completely inside (all 1's) or outside (all 0's) of \Pres and does not participate to the final surface. Vertex candidates may lay on the final boundary and still not participate to the final surface description %
if the vertex is introduced in the middle of a final edge. This happens as soon as the bit pattern of \v is symmetric along one of the components $i$ or $j$, which means that traversing the vertex along this border does not change the final primitive participation of the other polyhedron. A sufficient condition can thus be written as the predicate \texttt{isFinal2}$(\v)$, which rules out any topological symmetries along the $i$ or $j$ components, as underlined in subscripts: 
\begin{equation}
	\big((\b[\iemph{\ni}\nj], \b[\iemph{\ni}\yj]) \neq (\b[\iemph{\yi}\nj], \b[\iemph{\yi}\yj])\big) \wedge 
	\big((\b[\ni\iemph{\nj}], \b[\yi\iemph{\nj}]) \neq (\b[\ni\iemph{\yj}], \b[\yi\iemph{\yj}])\big)
\end{equation}
It can be noted that this condition includes the necessary conditions, since complete inclusion or exclusion is also a case of pattern symmetry. %

\textbf{Third order vertex.} At the intersection locus of three facets from three input polyhedra \P[\i], \P[\j], \P[\k], the vertex's neighborhood is locally split in eight subvolumes. The analysis is analogous to second order vertices, and requires examining the influence of crossing the three boundaries. The corresponding 8 combinations of \i, \j and \k bit-flippings are:
\begin{equation}
\begin{array}{rl}
	\b[\ni\nj\nk]&=\f(\I[1](\v), \cdots, 0, \cdots, 0, \cdots, 0, \cdots, \I[\N](\v))\\
	\b[\ni\nj\yk]&=\f(\I[1](\v), \cdots, 0, \cdots, 0, \cdots, 1, \cdots, \I[\N](\v))\\
	&\cdots \\ %
	\b[\yi\yj\yk]&=\f(\I[1](\v), \cdots, 1, \cdots, 1, \cdots, 1, \cdots, \I[\N](\v)).
\end{array}
\end{equation}
We call $\class=(\b[\ni\nj\nk], \cdots, \b[\yi\yj\yk]) \in \{0,1\}^8$ the classification vector of third order vertex \v.
Similarly to the order-2 case, complete inclusion or exclusion of the volume rules out the vertex, as well as any axis symmetries, which can be jointly evaluated with the predicate \texttt{isFinal3}$(\v)$: 
\begin{equation}
\begin{array}{rl}
	             &\big((\b[\iemph{\ni}\nj\nk], \b[\iemph{\ni}\nj\yk], \b[\iemph{\ni}\yj\nk], \b[\iemph{\ni}\yj\yk]\big) \neq \big(\b[\iemph{\yi}\nj\nk], \b[\iemph{\yi}\nj\yk], \b[\iemph{\yi}\yj\nk], \b[\iemph{\yi}\yj\yk])\big) \\
	\wedge &\big((\b[\ni\iemph{\nj}\nk], \b[\ni\iemph{\nj}\yk], \b[\yi\iemph{\nj}\nk], \b[\yi\iemph{\nj}\yk]\big) \neq \big(\b[\ni\iemph{\yj}\nk], \b[\ni\iemph{\yj}\yk], \b[\yi\iemph{\yj}\nk], \b[\yi\iemph{\yj}\yk])\big) \\
	\wedge &\big((\b[\ni\nj\iemph{\nk}], \b[\ni\yj\iemph{\nk}], \b[\yi\nj\iemph{\nk}], \b[\yi\yj\iemph{\nk}]\big) \neq \big(\b[\ni\nj\iemph{\yk}], \b[\ni\yj\iemph{\yk}], \b[\yi\nj\iemph{\yk}], \b[\yi\yj\iemph{\yk}])\big)
\end{array}
\end{equation}
If none of these conditions are met, the vertex is completely inside or outside the result polyhedron. If one (resp. two) of the conditions are met, the vertex is on a facet (resp. edge) of the final polyhedron.

{\bf Specificity of the 2-polyhedron case.}
Interestingly, in this situation, there are only first and second order vertices with no axis symmetries, \emph{i.e.} all order two vertex candidates participate to \Pres, which has been analyzed by~\cite{franco:hal-00904661}.%

\subsection{Vertex Retrieval Summary}

\newsymbol{\dir}{d}
\newsymbol{\face}{F}
\newcommand{\var}[1]{\texttt{#1}}
\newcommand{\Ifshort}[1]{\LState {\bf if }{#1 }{\bf then}}
\newcommand{\Elsifshort}[1]{\LState {\bf elsif }{#1 }{\bf then}}

We illustrate in \fig{csgvertices} how the set of \Pres's vertices may be retrieved using a simple but quartic worst-case complexity algorithm which loops over all 1, 2 and 3-facet combinations, using the previously defined \textsc{isFinal1}, \textsc{isFinal2}, and \textsc{isFinal3} predicates.  These predicates include rayshooting operations to compute the vertex's indicator vector, a linear operation with all input facets.
The algorithm uses classical intersection functions~\cite{geomtools4compgraph}:  \textsc{intersect2facets} computes the intersected edge between two convex facets, as a pair of vertices giving the edge extremities, and \textsc{intersectSegmentFacet} computes the vertex representing the intersection of  a segment and a facet, if any. Note that the quartic behaviour is in practice mitigated by the fact that the third order loop is only executed if a pair of intersecting input faces was already encountered.

\begin{figure*}
\begin{algorithmic}
\Function{CSGVertices}{}
\LState 
  {\bf Input:} \V*, \F*: set of vertices and facets of input polyhedra
\LState 
  {\bf Output:} $\V*[f]$: corresponding set of final output vertices
\vspace{-0.85mm}
\For{$\face[1]$  {\bf in} \F*}  \Comment{Enumerate input faces}

  \For{$\v $ {\bf in} $\face[1]$}   \Comment{Order-1 candidates}
     \Ifshort{\textsc{isFinal1}$(v)$}
        $\V*[f] \leftarrow \V*[f] \cup \{\v\}$
        \vspace{-0.85mm}
  \EndFor
  \For{$\face[2]$ {\bf in} \F*}
     \LState $\v[1],\!\v[2] \!\leftarrow$\! \textsc{intersect2facets}($\face[1]$,$\face[2]$) \Comment{Order-2 candidates}
	\Ifshort{$\{\v[1], \v[2]\}=\emptyset$} continue $\face[2]$ loop      \Comment{No intersection}
     \Ifshort{\textsc{isFinal2}$(\v[1])$} 
		 $\V*[f] \leftarrow \V*[f] \cup \{\v[1]\}$ 
		 \vspace{-0.90mm}
	\Ifshort{\textsc{isFinal2}$(\v[2])$}
		 $\V*[f] \leftarrow \V*[f] \cup \{\v[2]\}$
		 \vspace{-0.90mm}
     \For{$\face[3]$ {\bf in} \F*} \Comment{Order-3 candidates}
         \LState $v \leftarrow$ \textsc{intersectSegmentFacet}($\v[1]$, $\v[2]$, $\face[3]$)
         \Ifshort{$v \neq \emptyset$ \textbf{and} \textsc{isFinal3}$(v)$}
                  $\V*[f] \leftarrow \V*[f] \cup \{\v\}$
                  \vspace{-0.90mm}
     \EndFor
  \EndFor 
\EndFor
\EndFunction
\end{algorithmic}

\caption{Brute-force algorithm to find all vertices of the result polyhedron.
}\label{fig:csgvertices}
\end{figure*}

\section{Final Polyhedron Connectivity}
\secl{finalfacets}

\newsymbol{\Vres}{\mathcal{V}}[\f]
\newsymbol{\Fres}{\mathcal{F}}[\f]

Once the subset of final vertices \Vres is known through the  classification process, the main task left is to identify how vertices are connected together to form the  faces \Fres of the final polyhedron~\Pres. Our method makes a clear distinction between computing all geometric coordinates of final vertices and building the final topology. While the former involves numerical coordinate construction, the latter relies only on orientation and ordering predicates.

The classification vectors  introduced previously not only inform us of the participation of a given vertex \v to \Pres, they also give a  snapshot of volume and surface adjacencies around the vertex, as illustrated in \fig{geom}. We show here how to find all the polygons \v participates in, by introducing the \emph{looplet} construct. We define a looplet of \v as a loop fragment running through this vertex, represented by a symbolic incoming and outgoing edge direction. Its purpose is to compactly represent all partial adjacency information available for individual unconnected vertices, while being independently computable at each vertex.

\subsection{Surface and Edge Orientation}

Surfaces and edges  need to be oriented to define looplets. Surface boundaries contributing to the final result \Pres may change orientation (reversed normal), \emph{e.g.} when participating in a boolean subtraction. For a facet \face[a] of any polyhedron, we note the orientation of its contributions to the final surface as either $\face[a]<+>$ if the contribution conserves initial orientation, and $\face[a]<->$ if the orientation of the contribution is inverted.
In similar spirit we need to define an intrinsic edge orientation $\dir[ab]$ for every edge adjacent to two faces \face[a] and \face[b]. For this purpose we distinguish two cases:
\begin{itemize}
\item 
	if the edge is pre-existing  from an input \P[i], $\dir[ab]$ is the edge direction with polygon \face[a] on its left and \face[b] on its right on the oriented surface.
\item 
	if the edge arises from the intersection of \face[a] and \face[b], we define the edge direction $\dir[ab] = \face[a] \times \face[b]$, as the crossproduct of the corresponding face normals.
\end{itemize}
In both cases $\dir[ab] = -\dir[ba]$. With these definitions we can introduce a concise notation for looplets, as $(\dir[ab], v, \dir[ac] | \face[a]<+>)$, a looplet characterized as a positive contribution in \face[a], centered on vertex $v$, with incoming edge direction \dir[ab] and outward edge direction \dir[ac]. Looplets can be symbolically and compactly stored as a vertex reference for $v$, an orientation bit, and three ordered face references ($a$,$b$,$c$ in the above example) since the incoming and outgoing directions share the face $a$ in their adjacencies. In the following, we will describe how classification vectors  determine which looplets are present at a vertex, upon which the final polyhedron facets can be built. For this purpose, we break down the presentation of looplet generation cases for each vertex order. This breakdown is illustrated in the right column of \fig{geom}, where each subfigure shows in green the bit classification state deciding the presence of corresponding looplets, annotated under each figure in blue. %

\subsection{First Order Vertex Looplets}

A first order vertex \v that passed the classification test is on the surface boundary of a polyhedron~\P and also on the boundary of~\Pres. Each of its adjacent facets thus at least partially participates to~\Pres, and contributes a looplet for this vertex. The directions of this looplet are given by the edges adjacent to the facets of the looplet. If there is no change of orientation at the vertex \v, \emph{i.e.} $\b[\ni]=0$ and $\b[\yi]=1$, %
the edges and facet keep the orientation they had on the initial polyhedron, \emph{e.g.} yielding a looplet $(\dir[13],\v,\dir[12] | \face[1]<+> )$ for facet \face[1] in \fig{geom}. In contrast, looplets are inverted in case of surface orientation change, \emph{i.e.} when $\b[\ni] = 1$ and $\b[\yi] = 0$, \emph{e.g.} yielding looplet $(\dir[21],\v,\dir[31] | \face[1]<-> )$ for facet \face[1].

\subsection{Second Order Vertex Looplets}

Second order vertices are the intersection of the edge of a polyhedron \P[i] and the facet of a polyhedron \P[j]. As such it involves three facets, one facet labeled \face[2] from \P[j], and two facets \face[1] and \face[3] from \P[i], adjacent to the edge yielding \v from intersection with facet \face[2]. We here assume facet orientation as in \fig{geom}, without loss of generality: if the normal of one of the two surfaces is inverted, swapping $F_1$ and~$F_3$ brings us back to this reference configuration.

The crossing surfaces at the vertex define four boundaries between four subvolumes, each with two possible orientations. In the case of \face[2], each of the two subvolume boundaries may yield one looplet in each orientation, e.g. for the top subvolume boundary in \fig{geom} the two possible looplets are $(\dir[12],\v,\dir[32] | \face[2]<+> )$ and $(\dir[23],\v,\dir[21] | \face[2]<-> )$. One of the two possible looplets is generated for a subvolume boundary, as soon as the classification bits of the two subvolumes it separates have different values. The  looplet generated is the one consistent with the orientation of the final surface, with positive normal going from the inside ($\b[**]=1$) to the outside ($\b[**]=0$) of the final volume.

The decision scheme is analogous for looplets of \face[1] and \face[3]. Looplets of these facets are always simultaneously decided as they are determined by the same classification bits.

\subsection{Third Order Vertex Looplets}

\newsymbol{\t}{t}

The eight subvolumes surrounding \v are separated by the three facets \face[1], \face[2], \face[3].  The configuration shown in \fig{geom} assumes that the normals of these three facets form a right-handed trihedron. This is again without loss of generality, should one of the facets have an opposite normal, a single permutation in the order in which facets are considered brings us back to this reference configuration. Because the looplet possibilities are analogous for all three facet planes, we shall only enumerate the configurations for \face[1]. The enumeration also has a rotational symmetry within the facet plane, since it is divided in four quadrants by the other two facets. The four quadrants are indexed with $(r,s)\in\{0,1\}^2$ tuples. 
\fig{geom} shows that there are four possible looplets for a quadrant, bringing the total possible order-three looplet count to $3\times 4\times 4=48$ for a vertex. We focus our description on one quadrant of \face[1] where $(r,s) = (0,0)$.
To ease the description, we introduce two intermediate boolean predicates \t[rs]<+>, \t[rs]<-> for each quadrant $(r,s)$ of~\face[1]. They are computed as $\t[rs]<+>:=((\b[0rs],\b[1rs])=(0,1))$ and $\t[rs]<->:=((\b[0rs],\b[1rs])=(1,0))$, indicating whether the surface portion corresponding to the quadrant participates to the final surface as a positive or negative contribution in the plane of \face[1].

Looplet decisions involve examining several quadrant boundary predicates. Concave looplets exist if three of the four quadrant boundaries exist for an orientation and the fourth doesn't, \emph{i.e.} $(\dir[13],\v,\dir[12] | \face[1]<+> )$ exists if $((\t[00]<+>, \t[01]<+>, \t[10]<+>, \t[11]<+>) = (0,1,1,1))$, and $(\dir[21],\v,\dir[31] | \face[1]<-> )$ exists if $((\t[00]<->, \t[01]<->, \t[10]<->, \t[11]<->) = (0,1,1,1))$. On the other hand, convex looplets conditions depend only on three quadrant boundaries, the diagonally opposite quadrant in the facet having no influence.
Concerning quadrant $(0,0)$ in \fig{geom}, convex looplet $(\dir[21],\v,\dir[31] | \face[1]<+> )$ exists if $((\t[00]<+>, \t[01]<+>, \t[10]<+>) = (1,0,0))$, while the opposing looplet $(\dir[13],\v,\dir[12] | \face[1]<-> )$ exists if $((\t[00]<->, \t[01]<->, \t[10]<->) = (1,0,0))$.

\subsection{Retrieving Final Polyhedron Facets}

\begin{figure}
\begin{center}
\newcommand{\Ff}[1]{$F_#1$}
\newcommand{\Fv}[1]{$v_#1$}
\renewcommand{\Fd}[1]{$d_{#1}$}
\def\svgwidth{5cm}
\raisebox{-0.5\height}{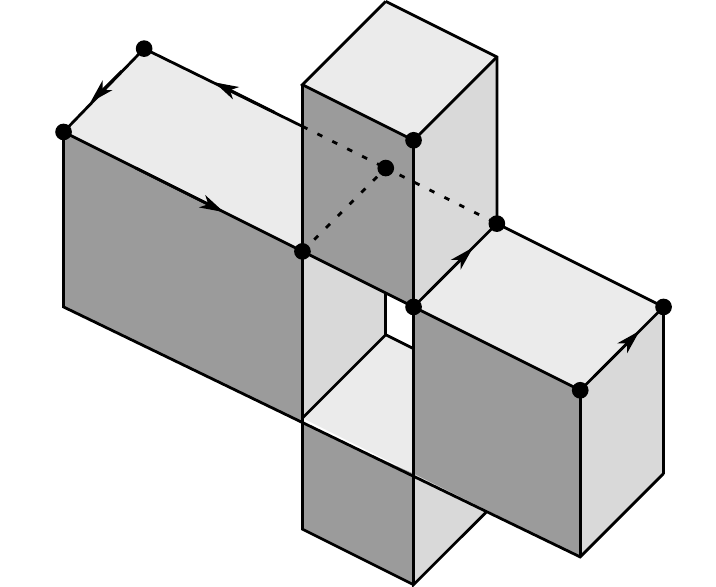}
\begin{tabular}{|c|}
\hline 
looplets\\
\hline
$(d_{06},v_{0},d_{01}|F_0^+)$ \\
$(d_{04},v_{1},d_{06}|F_0^+)$ \\
$(d_{05},v_{2},d_{04}|F_0^+)$ \\
$(d_{01},v_{3},d_{05}|F_0^+)$ \\
$(d_{30},v_{4},d_{01}|F_0^+)$ \\
$(d_{04},v_{5},d_{30}|F_0^+)$ \\
$(d_{02},v_{6},d_{04}|F_0^+)$ \\
$(d_{01},v_{7},d_{02}|F_0^+)$ \\
\hline
$(d_{05},v_{2},d_{40}|F_0^-)$ \\
$(d_{10},v_{3},d_{05}|F_0^-)$ \\
$(d_{03},v_{4},d_{10}|F_0^-)$ \\
$(d_{40},v_{5},d_{03}|F_0^-)$ \\
\hline
\end{tabular}
\end{center}
\caption{\label{fig:runningresult}
  Result of the $(\mathcal{P}_1 \mathrm{ xor } \mathcal{P}_2)\cap\mathcal{P}_3$ operation on the example of Figure~\ref{fig:runningexample}. The table lists the looplets for facet $F_0$.
}
\end{figure}

\newcommand{\ar}[1]{\overset{d_{#1}}{\longrightarrow}}

Once all vertices and their looplets have been generated, they can be re-indexed for  each facet to generate its contributions to \Pres. We process each input facet separately, which improves the locality of the algorithm and reduces it to 2D.
The corresponding algorithm is in~\fig{csgfacets} and an example of a result illustrated in \fig{runningresult}, where the focus is on contributions of~\face[0]. Within a finally contributing facet, an arbitrary seed looplet is chosen and its outgoing direction followed, searching for sequentially matching looplets to close the loop. In some occurrences, two or more looplets may match for a given direction: see for instance looplet $(\dir[06],\v[0],\dir[01] | \face[0]<+> )$ in \fig{runningresult}, for which both $(\dir[01],\v[3],\dir[05] | \face[0]<+> )$ and $(\dir[01],\v[7],\dir[02] | \face[0]<+> )$ match. In this case, the  \textsc{First} function selects the closest looplet in the search direction $d_{01}$, here $(\dir[01],\v[3],\dir[05] | \face[0]<+> )$. The whole process may be repeated until there are no looplets left in the face. For $F_0$, this results in the following loops: 
\[ \begin{array}{ll}
F_0^+:& v_0 \ar{01} v_3 \ar{05} v_2 \ar{04} v_1 \ar{06} v_0 \\ 
F_0^+:& v_4 \ar{01} v_7 \ar{02} v_6 \ar{04} v_5 \ar{30} v_4 \\
F_0^- :& v_4 \ar{10} v_3 \ar{05} v_2 \ar{40} v_5 \ar{30} v_4 \\
\end{array} \]
Two convex, diagonally opposing looplets may be triggered for a same vertex, as is the case for \emph{e.g.} \v[3] or \v[4] in \F[1]. Both negative and positive orientation facets may be generated for a single input facet, and may even be adjacent and share an edge, as for \face[0] in our example. Both of these configurations are typical of exclusive-or operations, but may happen with other operations. More generally, the algorithm can generate arbitrary output facets, with non-convex loops, several loops per facet (holes). Remarkably, our vertex-centered framework transparently accounts for all such possibilities.

Since facets may be arbitrarily large (see \emph{e.g.} Figure~\ref{fig:showcase} or Dithering in Figure~\ref{fig:results}), the cost of \textsc{First} searches may be quadratic if implemented naively. To avoid this, output vertices can be sorted lexicographically by vertex index and offset on the edge, to obtain quasilinear searching, in $\mathcal{O}(r\log r)$ with $r$ the number of vertices in the output facet.
At this stage the algorithm produced no superfluous geometry: all vertices and edges are required to represent the output polyhedron. However, non-convex and non-0 genus polygons are hard to manipulate, render or even feed back as input to the algorithm.
Therefore, we typically tesselate the output facets to triangles or convex polygons, also an $\mathcal{O}(r\log r)$ operation. This only happen once at finalization and never at an intermediate stage.

\newsymbol{\l}{l}
\newsymbol{\L}{\mathcal{L}}

\begin{figure*}
\begin{algorithmic}
\Function{CSGFacets}{}
\LState 
  {\bf Input:} \V*[f]: set of vertices of the output %
    \vspace{-.80mm}
\LState 
  {\bf Output:} $\L$: final polyhedron facets as set of loops
    \vspace{-.80mm}
	\LState $\F* \leftarrow \emptyset$, $\V* \leftarrow \{\}$ \Comment{Set of contributing facets, and their vertices}
  \For{$\v$ {\bf in} \V*[f]} \Comment{Collect looplets for all vertices}
    \vspace{-.80mm}
  		  \For{$\face$ {\bf in} \textsc{AdjacentFacets}$(\v)$}
  		  		\LState $\F* \leftarrow \F* \cup \{\face<+>\} \cup \{\face<->\}$ \Comment{Keep both orientations of facets}
				\LState $\V* [\face]$ $\leftarrow$ $\V* [\face]$ $\cup$ $\{\v\}$ \Comment{Facet vertex contributions}
		\EndFor
  \EndFor
  
  \For{$F$ {\bf in} $\F*$}\Comment{Process each facet's two orientations} 
  	\LState $\l \leftarrow \emptyset$ \Comment{Looplets indexed by incoming direction}
  	  \For{$\v$ {\bf in} $\V* [F]$} \Comment{Collect looplets for all vertices of $F$}
	  	\LState $\l \leftarrow \l \cup \textsc{ComputeLooplets}(*, \v, * | F)$
	\EndFor
  \While{ $\l \neq \emptyset$} \Comment{ Looplets left for this facet}
  		\LState $(\dir[1],v,\dir|F)$ = pop$(l)$ \Comment{ Pick and remove a looplet}
  		\LState  $F'\leftarrow []$ \Comment{ Build this final facet}
	    \Repeat \Comment{ Chain looplet vertices}
		    \LState $F' \leftarrow [F', \v]$
	    	\LState $(\dir',v,\dir| F) \leftarrow $ \textsc{First}$( \, (\dir,*,*| F)$ {\bf in} $\l \, )$
  		\Until{ $\dir = \dir[1]$} \Comment{Until back to start}
		\LState $\L \leftarrow \L$ $\cup F'$  \Comment{ Add this facet to final set}
  \EndWhile
  \EndFor
\EndFunction
\end{algorithmic}

\caption{Algorithm to find all facets of the result polyhedron from looplets.
}\label{fig:csgfacets}
\end{figure*}

\newsymbol{\Nf}{m}[]
\newsymbol{\h}{h}

\section{Hierarchical Algorithm}
\secl{kdtree}

A fully functional approach can be implemented based on the two simple algorithms, \textsc{CSGVertices} and \textsc{CSGFacets}. \textsc{CSGFacets} needs little tuning as it already runs quasilinearly over the output primitives identified by \textsc{CSGVertices}. However as previously noted the naive \textsc{CSGVertices} has an impractical quartic worst-case run time. We thus rely on a hierarchical space exploration to trigger \textsc{CSGVertices} only on tightly bounded space
regions where output geometry is expected to be found. We choose a KD-tree based exploration for this purpose because of its simplicity, and its favorable reported performance compared to other datastructures for similar tasks~\cite{havran00}.

\newsymbol{\c}{c}

\begin{figure}
\begin{center}
\includegraphics[width=\linewidth]{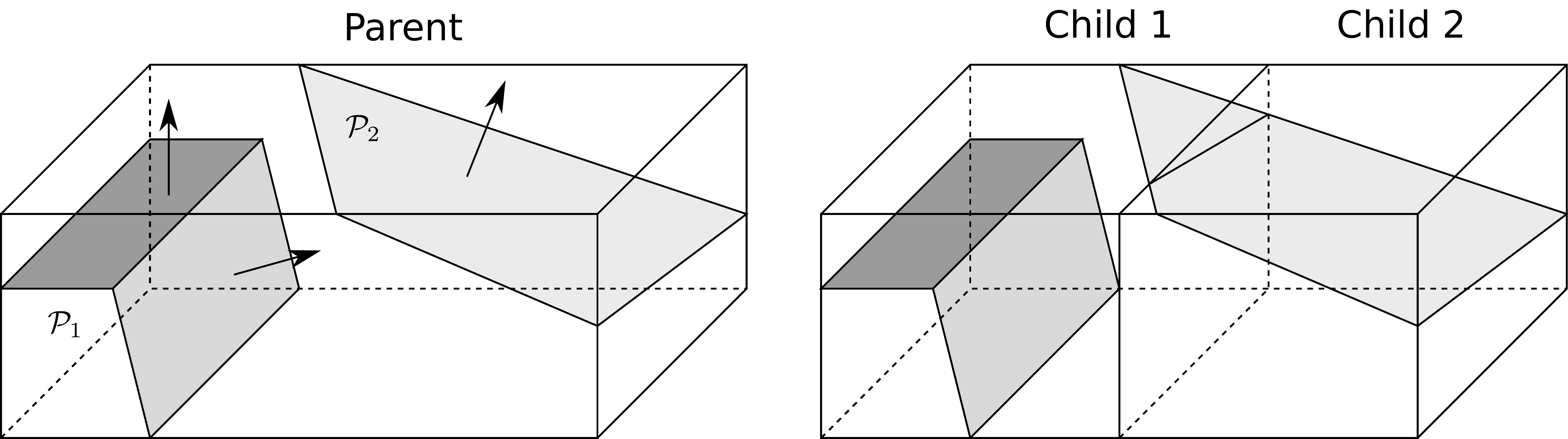}  \\
\begin{tabular}{|l||c|cc|}
\hline
node & parent & child 1 & child 2 \\
indicator vector & $(u,u)$    &  $(u,u)$     & $(0,u)$ \\
\hline
\end{tabular}
\end{center}
\caption{\label{fig:splitkdtreecell}
A KD-tree cell containing facets from $\mathcal{P}_1$ and $\mathcal{P}_2$ is split. The polygons stored in the KD-tree cells are cropped to the cell. 
}
\end{figure}

The KD-tree~\cite{bentley75} is a binary space partition tree, whose nodes each represent a particular axis-aligned cuboid cell, containing all cells of its child nodes. It is typically built as a search datastructure for a set of points in a d-dimensional space, by recursively subdividing the input point set in two subsets of equal size, using an axis-aligned split plane, achieving build complexities of $\mathcal{O}(m \log m)$ time and $\mathcal{O}(m)$ space with $m$ input points~\cite{berg08}. As applied to polyhedra, the KD-tree cells contain a list of polygons, cropped to the cell's bounding box. When polygons sit across a split plane, they are divided in two fragments inherited by both children of the parent node (\fig{splitkdtreecell}). Typically the cell subdivision is pursued down to a certain depth or until the number of primitives in the cell falls under a chosen bound, below which brute force search is more efficient.

\subsection{KD-Tree Exploration for the Boolean Problem}

\newcommand{\nodeB}[1]{\protect\raisebox{0.25em}{\fcolorbox[rgb]{0,0,0}{#1}{ }}}

\begin{figure*}
\begin{center}
\includegraphics[width=\linewidth]{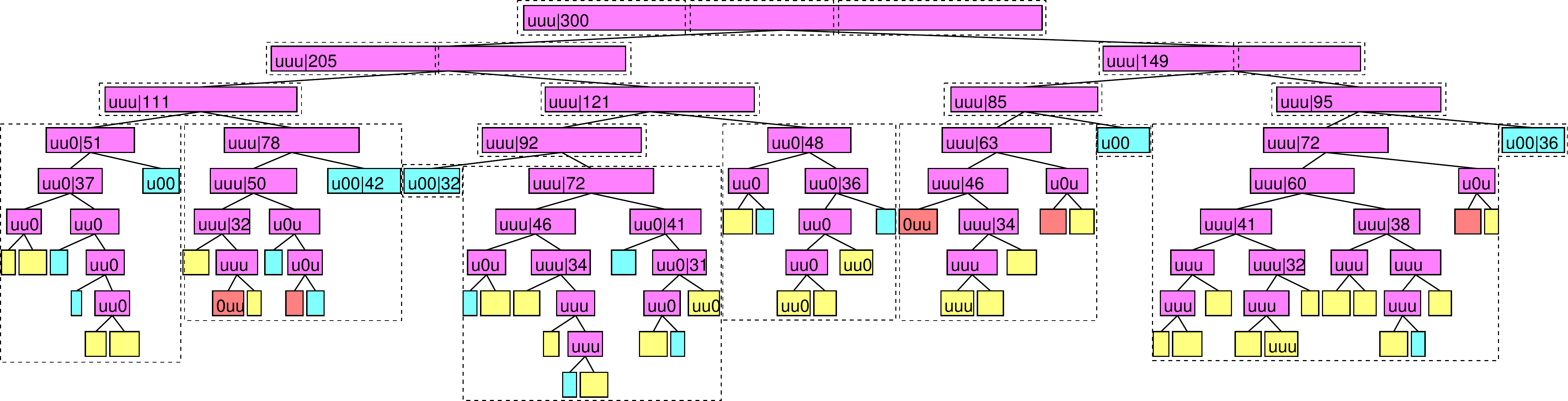}
\end{center}
\caption{\label{fig:kdtree}
  View of the KD-tree for operation $\mathcal{P}_f = \P[1] \backslash (\P[2] \cup \P[3])$ applied to three simple meshes. Each box represents a node, whose width is proportional to its number of polygons. When space allows, text in the box indicates node's indicator vector and the number of polygons. Color code: \nodeB{1,0.5,1} = node that was split, \nodeB{1,1,0.5} = leaf where vertices were found with~\textsc{CSGVertices}, \nodeB{0.5,1,1} = facets were just copied to the output, \nodeB{1,0.5,0.5} = node was found completely inside or outside the mesh. Each dashed box represents a parallel task. 
}
\end{figure*}

\begin{figure*}
\begin{center}
\makebox[0pt]{
\includegraphics[width=\linewidth]{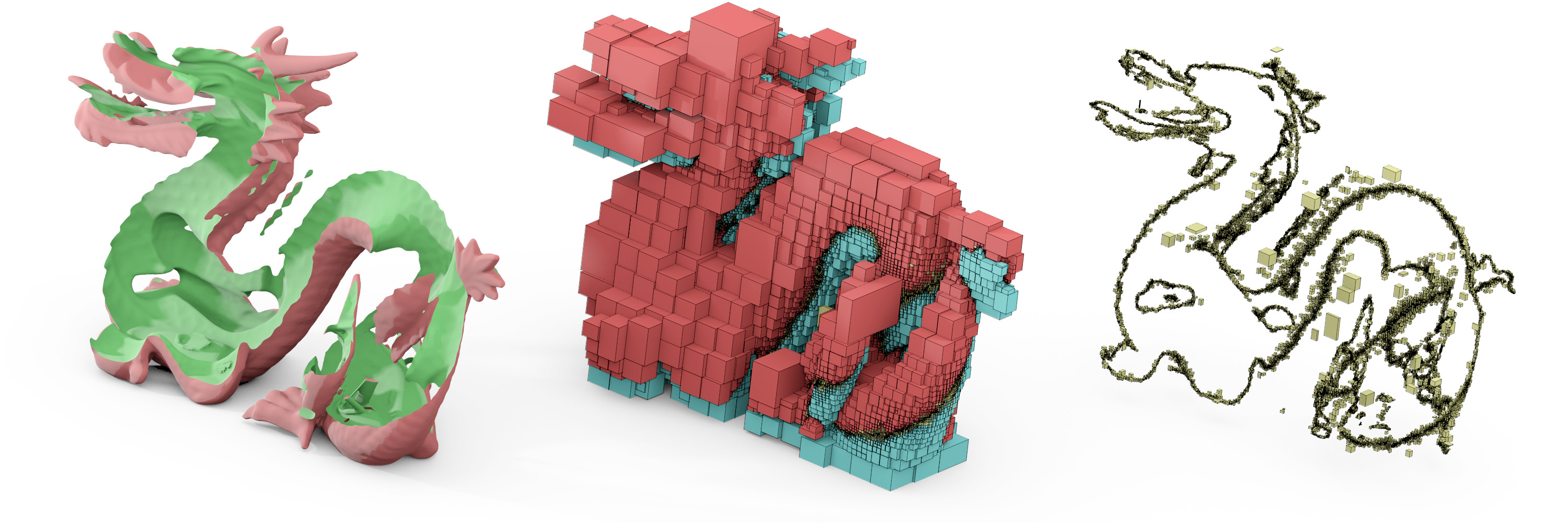}
}
\vspace*{-2ex}
\end{center}
\caption{\label{fig:kdtreedragon}
	Left: Boolean difference between a red and a green dragon mesh, center: bounding boxes of leaf nodes of the KD-tree, right: the leaves where \textsc{CSGVertices} is called (on average, 0.56 order-2 vertices are found on each of these). Color code, center and right: %
 \nodeB{1,1,0.5} = leaf where vertices were found with~\textsc{CSGVertices}, \nodeB{0.5,1,1} = facets were just copied to the output, \nodeB{1,0.5,0.5} = node was found completely inside or outside the mesh. The yellow boxes are barely visible in the central representation because the vast majority of output does not require intersection computations.
 }
\end{figure*}

Our procedure is similar to KD-tree construction, with the key differences that we never need to store the tree, and that our tree exploration is adaptive and output-sensitive. We want to subdivide only the nodes that may contain order-2 or -3 vertices of the final polyhedron, while pruning others, see Figure~\ref{fig:kdtreedragon}. This requires classifying KD-nodes during construction for every polyhedron involved. A volumetric cell can be classified as being fully inside or outside a polyhedron, but it can also straddle the polyhedron surface as soon as it contains surface primitives, and be thus undecided. We therefore need to extend the binary boolean logic to ternary as in \cite{pavic2010hybrid}. We define three corresponding indicator states \I[i](\c) for a cell \c with respect to polyhedron \P: $\I[i](\c) \in \{0, 1, u\}$ and the cell's indicator vector $\Iv(c) = (\I[1](\c), ..., \I[\N](\c))$ as the tuple of the cell's ternary indicators. In turn, final classifications $\Ires(\c) = \f(\Iv(\c))$ of a cell \c are computed with ternary logic, with:
\begin{equation}
f: \{0,1,u\}^{\N} \rightarrow \{0,1,u\}.
\end{equation}
Although the indicator vector of the cell may contain undefined bits, it is often possible to get a definite answer for the cell classification with \f. For example, consider the intersection of \N polyhedra, where $\f[\cap](\Iv(c))=1$ iff $\Iv(c)=(1,1,\cdots,1)$. %
It suffices for the cell to be known outside of any polyhedron to conclude that the cell is outside the intersection volume, \emph{i.e.} $\f[\cap]( \cdots, u, \cdots, 0, \cdots ) = 0$. Likewise for unions, $\f[\cup]( \cdots, u, \cdots, 1, \cdots ) = 1$, regardless of other bits, the cell is known to be inside the union as soon as it is in one of the input polyhedra. Truth tables for ternary versions of usual boolean operations are easy to build. Note that for specific hard functions such as \f[\mathrm{xor}], \emph{all bits} of \Iv(c) must be defined ($\neq u$) to compute a definite classification, in other words all order 2 and 3 vertices participate to the final surface. Nevertheless whatever the boolean function \f, it always benefits from other generic pruning features of the algorithm discussed below.

\subsection{Algorithm Summary}
\newsymbol{\fmax}{\mathrm{F}}[\mathrm{max}]

\begin{figure*}
\begin{algorithmic}
\Function{KDVertices}{}
\LState 
  {\bf Input:}  \V*, \F*, \Iv*: cell's input vertices, facets and indicator vector
\LState 
  {\bf Output:} $\V*[f]$: cell's output vertices
\vspace{-0.85mm}

\If{\f(\Iv*) $\neq u$} 
\Comment{Cell is completely inside or outside \Pres}
\vspace{-0.95mm}
\LState {\bf return} $\emptyset$
\Comment{Pruning, no surface primitive contained}
\ElsIf{$|$\textsc{Undefined(\Iv*)}$| = 1$} 
\Comment{Only one active input \P}
\vspace{-0.95mm}
\LState {\bf return} \V*
\Comment{Pruning, all cell vertices participate to \Pres}
\vspace{-0.90mm}
\ElsIf{$|\F*| \leq $ \fmax}
\Comment{Threshold cell size is reached}
\LState {\bf return} \textsc{CSGVertices}(\V*, \F*)
\Comment{Find cell's order 1,2,3 vertices}
\Else
\Comment{Cell may still have final vertices}
\LState \V*<1>, \F*<1>, \V*<2>, \F*<2> $\leftarrow$ \textsc{Split}(\V*, \F*)
\Comment{Choose and execute split}
\LState $\Iv*<1>, \Iv*<2> \leftarrow \Iv*$
\Comment{Subnode indicators are updated from \Iv*}
\For{\i  { \bf in} \textsc{Undefined(\Iv*)}}  \Comment{Enumerate still active \P's }
\vspace{-0.95mm}
\LState $\Iv*<1>[i] \leftarrow$ \textsc{isInside}(\V*<1>, \P)
\Comment{Update i-th bit}
\vspace{-0.95mm}
\LState $\Iv*<2>[i] \leftarrow$ \textsc{isInside}(\V*<2>, \P)
\vspace{-0.95mm}
\Comment{Update i-th bit}

\EndFor

\LState {\bf return} \textsc{KDVertices}(\V*<1>, \F*<1>, \Iv*<1>)  $\:\:\cup$ 
\textsc{KDVertices}(\V*<2>, \F*<2>, \Iv*<2>)
\EndIf
\EndFunction
\end{algorithmic}

\caption{Hierarchical algorithm to find all vertices of the result polyhedron.
}\label{fig:kdvertices}
\end{figure*}

The algorithm is summarized in \fig{kdvertices}. There are two pruning conditions. First, if the cell final classification \f(\Iv(\c))becomes completely determined, then it does not contain final surface primitives. Second, if the cell is still undetermined but only one polyhedron participates, then all of its primitives in the cell are part of the final polyhedron and no recursion is needed. This condition is expressed by counting the undefined bits of \Iv(\c), where \textsc{Undefined}(\Iv*) denotes the set of indices of polyhedra whose surface still runs through the cell. As evidenced by the example of \fig{kdtreedragon}, for large meshes, most of the explored nodes will fall in these two pruning cases, and the algorithm subdivisions can be seen as converging to the final surface's order-2 and order-3 intersections.

If the number of facets falls under a threshold \fmax, we fall back to \textsc{CSGVertices} to report final vertices. Experimentally $\fmax\!=20$ is found to be efficient in all situations. Specifics of local indicator evaluation in \textsc{CSGVertices} will be discussed in~\secref{leaf}.  Finally, if none of the previous conditions were met, the primitive set can be split, and the two subtrees explored once their indicator status has been updated. \textsc{isInside}(\V*<*>, \P) returns a value in $\{0, 1, u\}$: $0$ if all vertices of \V*<*> are outside \P, $1$ if they are all inside, $u$ otherwise. Since splitting is at the heart of the algorithm complexity, and \textsc{isInside} is performed jointly for the two children, we specifically address their implementation in \secref{nodesplit}.

\subsection{Computing Splits and Node Indicators}
\label{sec:nodesplit}

Various split heuristics exist and are widely documented in the litterature~\cite{havran00}. They determine the tree balancing and the number of split polygon fragments, which may clutter the tree and its performance. Yet the split must be performed as rapidly as possible to keep overall runtime under control. The complexity should be $O(m)$ with $m$ the number of primitives. Pre-sorting input vertices along the axis to compute medians without re-sorting each node, or minimizing polygon splits with the Surface Area Heuristic (SAH)~\cite{macdonald90}, are typical optimizations used to achieve $O(m \log m)$ tree build performance with small overhead for various tasks~\cite{wald06}. However, our objective is different, because entire subtrees are to be pruned and the split algorithm can hardly anticipate where. %
Minimizing the number of splits or finding the median of input polygon vertices is not as important to our algorithm as favoring large pruning possibilities. %
We have tried several advanced heuristics for pruning, but found experimentally that simply splitting in the middle of the bounding box's largest dimension yields excellent overall performance.

As soon as a the splitting plane is decided, a single $O(m)$ pass can build the split primitive sets. This pass can also be used to compute new bounding boxes for each sub-tree. Checking whether child nodes intersect the $i$-th bounding box provides information to update  the $i$-th bit of the child node's indicator vector (which is denoted \textsc{isInside} in the algorithm of~\fig{kdvertices}). If it still intersects, the $i$-th bit stays undefined, $u$. If no longer active, the child node contains no $i$-th polygon fragment, but we still need to determine whether the cell is completely inside or outside the $i$-th polyhedron.
Shooting a ray outside the current cell would require examining all input primitives and downgrade performance. Fortunately it is possible to answer the question locally by keeping a reference to extremal vertices in the split direction during the splitting pass. %
Namely, we examine the facets adjacent to the extremal vertex, and if they do not have the same normal orientation, we select the facet closest (most parallel) to the splitting plane for the orientation decision.
In the example of
Fig.~\ref{fig:splitkdtreecell}, the normal of the $\mathcal{P}_1$ polygons in
child 1 are used to compute bit 1 of the indicator vector of child 2.

\subsection{Computing the Indicator Vector of Leaf Points}
\secl{leaf}

Recall that \textsc{CSGVertices} requires ray shooting to compute indicators of candidate vertices in the general algorithm. In the context of the KD-exploration, this rayshooting occurs for calls of \textsc{CSGVertices} in leaves, and can be made local to keep the computational time bounded. Given the indicator vector \Iv(\c) for a leaf node \c, the $i$-th bit of the indicator vector \I(x) can be computed as follows: 
\begin{itemize}
\item 
  if  $x$ is on a facet of polyhedron $\P$, then $\I(x) = s$ (surface bit), %
\item 
  if the cell indicator is known $\I(\c) \ne u$, all points inside inherit the indicator bit $\I(\x) = \I(\c)$,
\item
  otherwise $\I(\v)=u$, we consider the non-empty set $P$ of
  polygons of $\P$ in the node. We shoot a ray from $x$ to an
  arbitrary point of one of the polygons to ensure at
  least one  intersection with $\P$. Then we compute the intersection
  of this  ray with all polygons of $P$ an keep the intersection nearest to $x$. The sign
  of the dot product of the ray's direction with the normal at the
  closest point gives the indicator bit \I(\x).
\end{itemize}

\newcommand{\Ninleaf}{N_\mathrm{inleaf}}
\newcommand{\Csplit}{C_\mathrm{split}}
\newcommand{\ccheck}{C_\mathrm{check}}

\subsubsection{Jittering}

To avoid feeding the algorithm degenerate configurations, two workarounds are implemented: 
\begin{itemize}
\item
	for CAD meshes, it is useful to randomly translate each meshes independently by a random vector. The vector should be small enough not to change the topology of the output, but still the same order of magnitude as the input mesh size. The random translation is reverted, ie. the true intersection vertices are recomputed from the faces they are the intersection of.
\item	
	the KD-tree is axis-aligned, so axis-aligned facets may also produce degeneracies. A simple workaround is to apply a random rotation to the input and revert this rotation at the end. \end{itemize}

No explicit check is done that the random jitter does not introduce new degeneracy: maybe the random rotation aligns another facet with the bounding boxes? This is because the probability computation above shows that such coincidence is almost impossible.

\section{Parallel Implementation}
\secl{parallel}

\begin{figure*}
\newcommand{\igantt}[1]{\framebox{\includegraphics[scale=0.23]{figs/gantt/#1_nt48_idfreeze}}}
\begin{center}
\makebox[0pt]{
\begin{tabular}{c@{}c@{}c}
T1 (47 ms) & T2 (37 ms) & H (125 ms) \\
\igantt{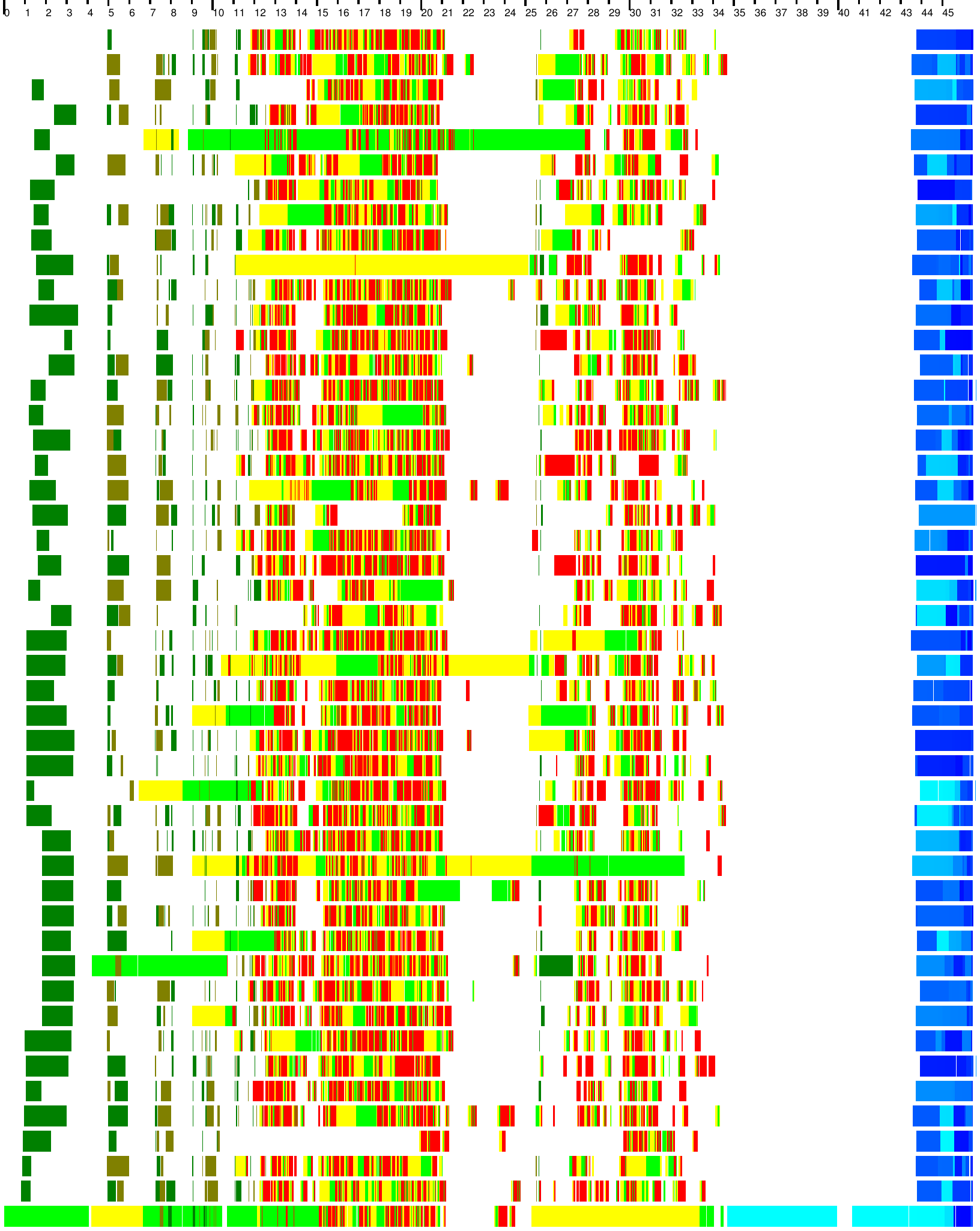} & \igantt{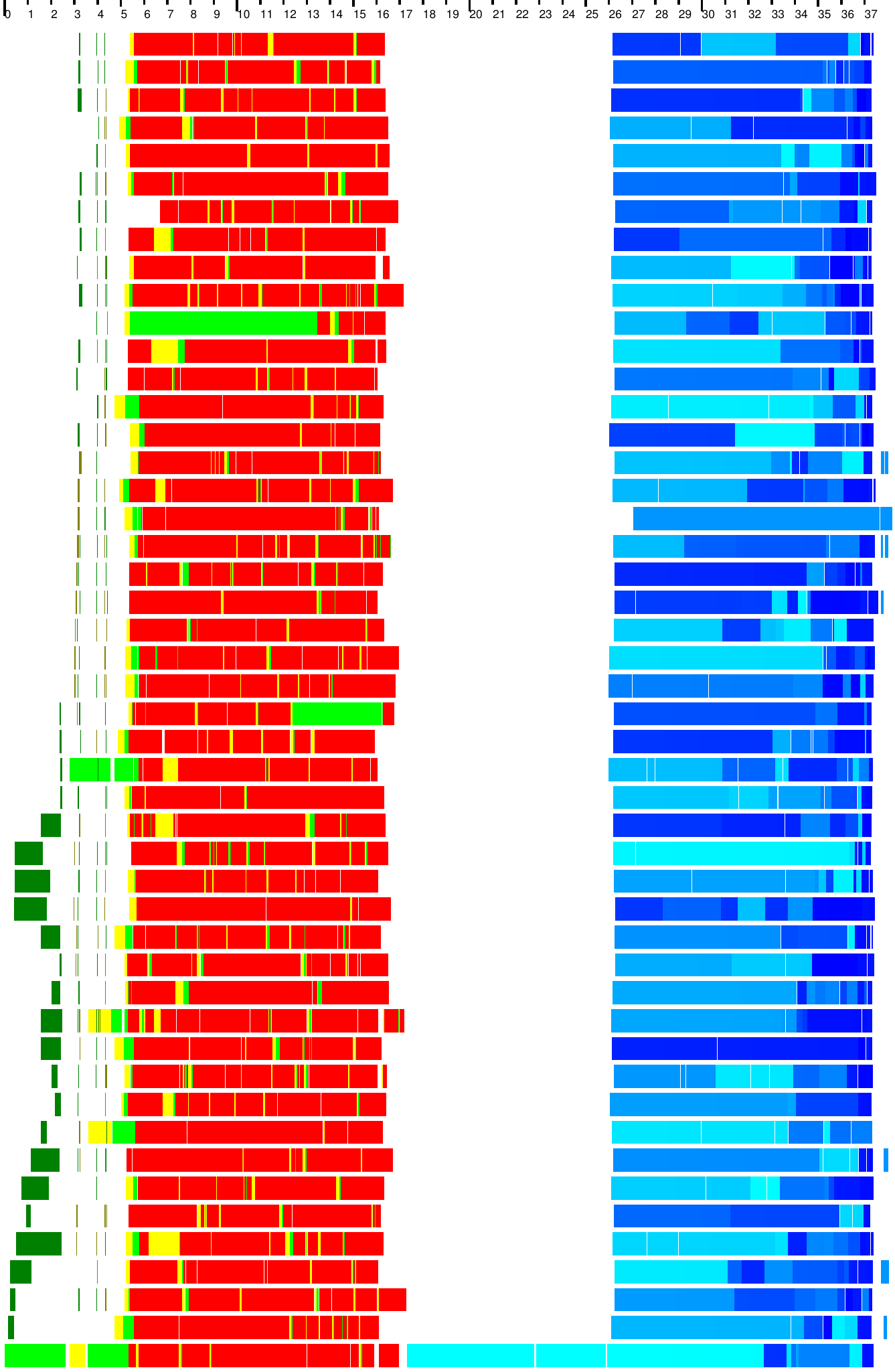} & \igantt{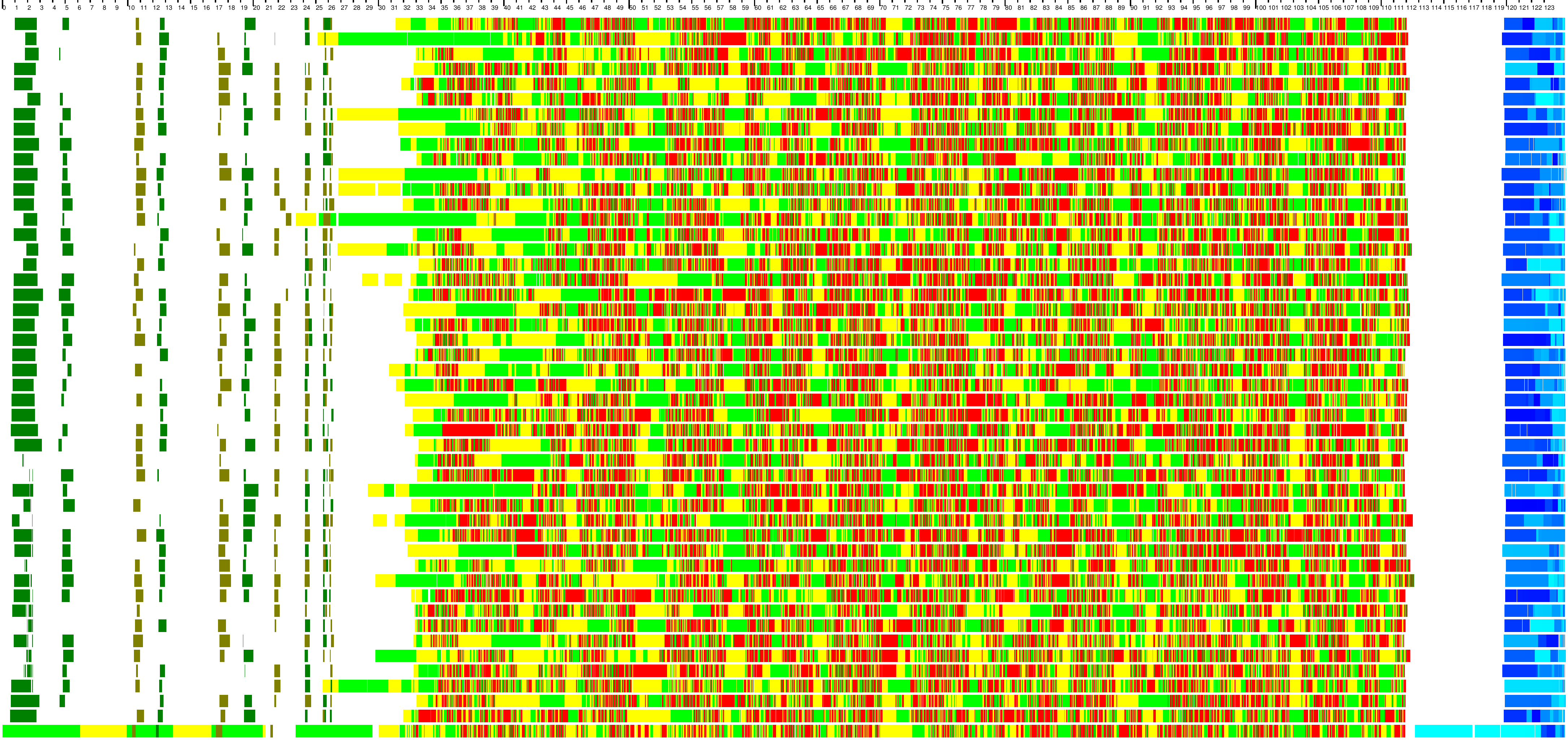} \\
\end{tabular}
}
\end{center}
\caption{\label{fig:gantt} 
Gantt chart of QuickCSG running on a 48-core AMD machine on 3 different
datasets (left:T1,central: T2,right: H). Datasets are detailed in
section~\ref{sec:datasets}. Each line coresponds to a core ativity
(white: idle).  Node splitting  is
 performed in the green and yellow bars, while  internal node
 parallelization is represented by dark green bars. Red bars are leaf
 computations (calls to \textsc{CSGVertices}). A  light blue sequential step
concatenates results from the different cores before building facets in
 parallel (calls to \textsc{CSGFacets}, shades of blue).  Notice
 that total execution times (indicated on top) are degraded due to code instrumentation
 to gather these statistics.}
\end{figure*}

Performance being a main concern, parallelization is mandatory to benefit from today's mainstream multi-core computer architectures. Parallelization has been studied in the context of boolean solid modeling, with elementary operations queued and balanced among processors~\cite{krishnan01}, by concurrently computing the result of non-dependent node operations in the CSG tree, or by using the GPU-friendly Layer Depth Image as CSG approximation~\cite{wang2011csg}. KD-tree algorithms have also been parallelized in the context of ray-tracing~\eg~\cite{Choi2010} and \cite{shevtsov2007highly}.
The biggest common issue algorithms face is the irregularity of the tree exploration, because work associated to each node of the KD-tree
is data-dependent and difficult to predict. A good workload balancing strategy is necessary for optimal resource and core usage. Instead of crafting very specific code, we use the work-stealing paradigm for tree exploration, for which off-the-shelf scheduling algorithms exist.

\subsection{Work Stealing Principle}

Work stealing frameworks allow to express the potential parallelism by delimiting dynamically created tasks that can be executed concurrently. 
Each processing core maintains a list of tasks.  When a core generates a task, it pushes it in its local list.   A task of this list is ready for execution once synchronization  constraints have been resolved. 
When a core becomes idle (\ie no local task left),  it randomly selects another core and steals
part of the tasks ready to be executed in the task list of its target. 
If no task can be stolen, an other victim is targeted. This scheduling
algorithm has  proven  performance~\cite{Blumofe:1999}.   Today, several parallel programming
environments are based  on work stealing (Cilk, TBB, OpenMP,
KAAPI). They come with high level constructions easing the parallelization of common patterns (loop with
independent iterations for instance).  Their  implementations ensures
high performance on  multi-core processors and shared memory machines.
We parallelized QuickCSG   with  Thread Building Blocks (TBB). 

\subsection{Proposed Implementation} 

The  recursive nature of the  KD-tree construction fits the task
model well. We encapsulate the \textsc{KDVertices} node processing function in a task. 
These tasks can be executed concurrently and  work stealing ensures 
they are dynamically spread amongst enrolled cores.

The KD-tree exploration starts  with a single  task.  
Enough tasks become available to keep all cores busy  only once a certain
depth is reached. Meanwhile,  many cores will stall. %
To circumvent this bottleneck,  \textsc{Split} calls in the toplevel nodes are parallelized internally, with a simple parallel for, %
with results  accumulated in  
separate vectors for each thread. Since this is   less efficient than the node-level parallelization,  this internal  parallelization  is  enabled only  in the very upper levels of the tree. 
Creating a task comes with some overhead, that can become significant for nodes  with a  light compute
load. This is the case for deep nodes  where the number of tasks is
much higher than the number of enrolled cores.
Thus to shave off overheads, we turn to a sequential sub-tree exploration
once the number of facets to process in a node is below a given
threshold (80). 
The results, spread in thread-local data structures, are then
sequentially concatenated  in a  global data structure before the call  to
the \textsc{CSGFacets} function, which is easily parallelized with a 
parallel for. 

Figure~\ref{fig:gantt} shows a gantt  chart for
executions of QuickCSG on three different datasets.  Though top node splitting  is
parallelized internally (dark green), it is not as efficient as the node level parallelization once enough nodes
have been generated (yellow, green and red bars). The sequential
concatenation of results (light blue bars) incurs a non-negligible cost at 48 cores.  This step is  very memory-intensive,
drastically limiting the efficiency of any parallelization. 
The scene geometry  greatly influences the execution. 
For instance  the number of output facets is much higher than the
input facets for T2 (central chart), which explains the relatively
high cost of  the calls to  \textsc{CSGFacets} (blue bars).

%

%
%

\section{Experiments}
\secl{results}

We implemented the QuickCSG  algorithm in C++, using 64-bit floats for coordinates. We rely on TBB for thread-stealing implementation, with optimized thread local memory allocations. The parallel implementation reverts to sequential exploration when there are less than 80 faces in the node, and brute-force \textsc{CSGVertices} is called as soon as the number of faces drops below 20. We store indicator vectors for nodes and points as 64-bit word bitfields. We use the particularly robust and efficient GLU implementation for the output polygon tesselation. 

\begin{figure}
\begin{center}
\includegraphics[height=3.5cm]{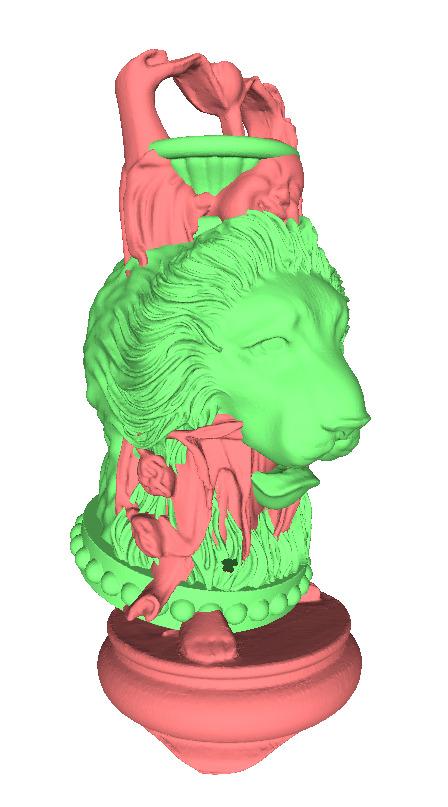}%
\makebox[0pt][l]{\textcolor{blue}{\raisebox{1cm}{\hspace*{-1.2cm}\fbox{\rule{8pt}{0pt}\rule{0pt}{3pt}}}}}
\hspace{5mm}
\includegraphics[height=3.5cm]{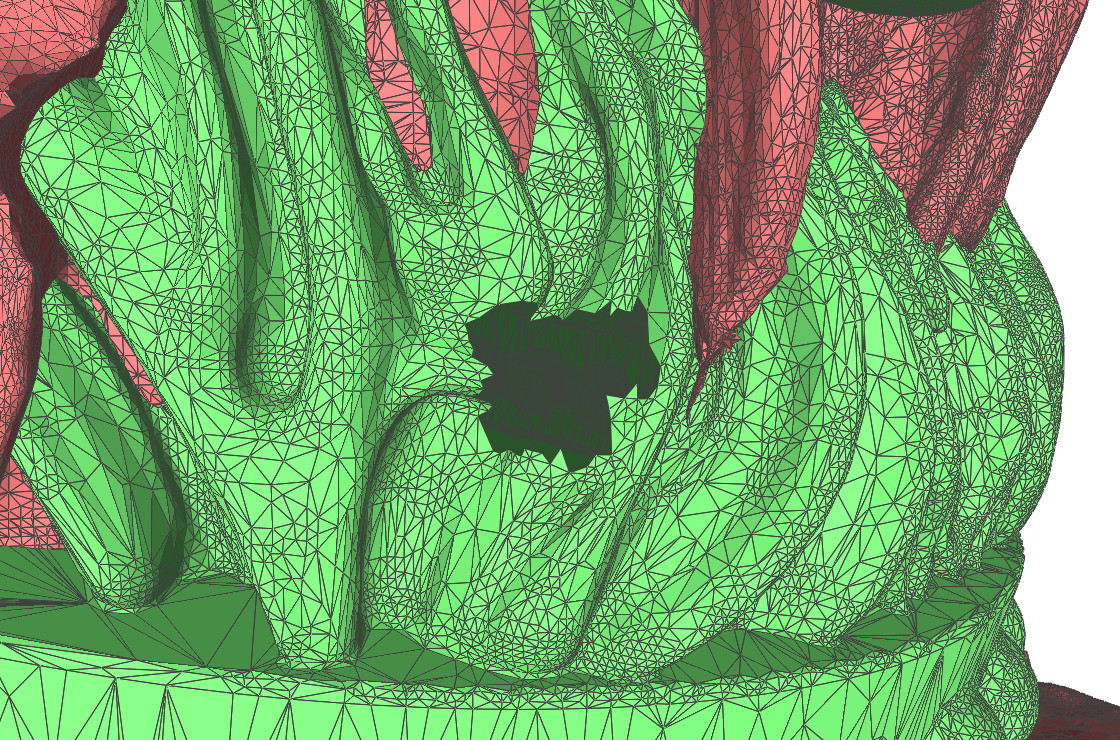}
\end{center}
\caption{\label{fig:failcase}
  Failure case: when computing the union between two meshes, the
  algorithm makes a mistake on the mesh position of a KD-tree node because the input mesh is invalid (self-intersecting). Left:
  the result, right: close-up of the hole in the mesh (the example is ``Buddha $\cup$ Vase-Lion'' from Section~\ref{sec:cmpmeshworks}).
}
\end{figure}

Similar to~\cite{wang2011csg,feito13}, we observed that many CSG implementations (\eg CGAL, 3DSMax ProBoolean and Carve CSG) crash, generate empty results or refuse to process meshes that they cannot handle completely. Non-robustness to near degeneracies, inability to cope with large and dense primitives, incompatibility of inputs with processing hypotheses are the typical causes. Violation of QuickCSG's input hypotheses may also occur for certain input meshes, \eg in \fig{failcase} the input polyhedra are self-intersecting, which affects \textsc{isInside}, leading to a hole in the bounding box of an incorrectly cancelled node.
Regarding numeric robustness, we resort to $\epsilon$-coordinate jittering, which, combined to the fact that only output primitives are computed, are observed to vastly reduce and in most cases eliminate the occurrence of degeneracy problems. In both cases QuickCSG  fails gracefully: it records the occurrence of errors while outputting the constructed result, even if partially incomplete, as in \fig{failcase}. In the experiments presented, we state when errors have occured. On QuickCSG's web page and the supplementary material, we provide the data and command lines that reproduce the experiments on QuickCSG.

In this section, we compare QuickCSG to state-of-the-art CSG
implementations on their own provided benchmarks. Because such benchmarks are typically limited to a few operations on medium to large size meshes, we introduce a new set benchmarks with more meshes and more complex CSG operations.We experimentally probe the main characteristics of QuickCSG: its complexity, parallel performance, and gains of $n$-ary versus binary operators. %

Unless stated otherwise, we ran the experiments on a i5 CPU 750 at 
2.7~GHz (4 cores) with 4~GB of RAM.  Reported   execution times
encompasses the processing from input meshes to the output mesh
including tesselization to convex polygons,  but excluding  startup
time and disk I/O.  The reported timings are wall-clock times in seconds, measured by
the \texttt{gettimeofday} function.  As  timings for several runs  are within 1\% of each other,
we do not report standard deviations.

\subsection{Comparison With State of the Art}

Our comparisons focus on recent boolean methods ~\cite{wang2011csg,wang2011csg,pavic2010hybrid,feito13}  and one
software package (Carve).  To the best of our knowledge these
implementations are currently the most computationally efficient ones. The related papers provide timing comparisons with other public (CGAL, GTS)~\cite{feito13} or
commercial packages (Rhino, ACIS~\cite{wang2011csg}, Houdini~\cite{pavic2010hybrid}, and 3DS Max~\cite{feito13}), which  were consistently found
to be significantly slower. We thus focus our comparisons on the former, with input data obtained either directly from the authors, or from the Stanford 3D scanning repository, 
resizing the meshes to the detail level reported in the papers if necessary. 
In most cases, the authors did not provide their implementation, so we directly
compare QuickCSG to the published execution times, with some comments about the differences between the hardware.  For these experiments, we use single-core sequential runs of QuickCSG, unless otherwise stated.

\subsubsection{Carve CSG}

Carve is the CSG library\footnote{Carve 1.4 can be found at
  \url{http://code.google.com/p/carve/downloads/list}.} used
in the Blender modeler. It uses an octree accelerator structure and produces 
clean output meshes, with few useless vertices.
We compare Carve and QuickCSG  on the most time consuming benchmarks
of the Carve CSG test suite (\texttt{test\_intersect.cpp}). %
QuickCSG is 4 to 14 times faster than Carve on these examples handling large meshes or/and many meshes:
\begin{center}
\resizebox{\columnwidth}{!}{%
\begin{tabular}{|ll|rr|}
\hline
Example & $\Nf$ & Carve & QuickCSG \\
\hline
21: sphere $-$ translated sphere   & 19600  & 0.342 &  0.077 \\
29: union of 30 rotated cubes      & 180    & 0.495 &  0.077  \\
30: sphere $-$ sphere $\cap$ cube  & 19606  & 0.469 &  0.032  \\
34: cow $\cup$ translated cow      & 185728 & 3.601 &  0.313  \\
\hline
\end{tabular}%
}
\end{center}

\subsubsection{MeshWorks}
\label{sec:cmpmeshworks}

MeshWorks\footnote{Executable
  at~\url{http://www2.mae.cuhk.edu.hk/~cwang/projMeshWorks.html}} is a GPU implementation 
  of an approximate Layered Depth Image algorithm~\cite{wang2011csg}. Because the intersection of surfaces is only resolved up to the resolution of the depth layer images used, several resolutions are tested in their paper. The range of timings in the following table reflect the range from coarse to fine resolution, with the finer scale taking the most time. It is the most relevant to our tests since QuickCSG does not perform any approximation and retrieves the exact topology.
QuickCSG, executed on 4 cores,   is from 10 to 40 times
faster  than MeshWorks executed on a nVidia GTX 260 GPU (+ 4 CPU cores),  
on the  examples provided along with the software:
\begin{center}
\resizebox{\columnwidth}{!}{%
\begin{tabular}{|ll|rr|}
\hline
Example & $\Nf$ & \cite{wang2011csg} & QuickCSG \\
\hline
Dragon $\cup$ Bunny     & 941k  & 55.4  & 3.4 \\
Small dragon $-$ Bunny        & 347k  & 3.06 -- 8.97  & 0.253\\
Buddha $\cup$ Vase-Lion & 1.48M & 10.68 -- 21.81 & 1.027\\
\hline
\end{tabular}%
}
\end{center}

We compare both with the provided MeshWorks implementation (first example) and timings from the original paper (two last examples).
Notice that the input meshes have small self-intersections which violates the input assumptions, so the QuickCSG output contains holes, see Figure~\ref{fig:failcase}.

MeshWorks is optimized for large and detailed meshes. In this case,
most faces do not intersect another face.  Once identified, these faces can be
directly copied to the output. MeshWorks and QuickCSG support this,
but it seems that the up- and down-load to the GPU hurts MeshWork's 
performance.

\subsubsection{Hybrid Booleans}

The ``Hybrid Booleans'' method of~\cite{pavic2010hybrid} subdivides
the input space with an octree, then constructs an approximate output mesh by remeshing the surface at the resolution of leaf nodes. 
We run QuickCSG on the paper's test data, and compare it
against the reported timings, as the implementation is not
available. %
Depending on the quality settings for~\cite{pavic2010hybrid}, QuickCSG
is 5 to more than 100 times faster. Again the higher times are more relevant to the comparison as QuickCSG retrieves exact topology on all these examples:

\begin{center}
\begin{tabular}{|ll|rr|}
\hline
Example & $\Nf$ & \cite{pavic2010hybrid} & QuickCSG \\
\hline
Chair & 1.5k & 1.3 -- 13 & 0.003 \\
Sprocket & 11k  & 5 -- 47  & 0.069 \\
Organic & 219k & 1.6 -- 24 (+1) & 0.488 \\
\hline
\end{tabular}
\end{center}

The  Hybrid Booleans method  requires to 
explore the tree to a predefined depth even for the simplest of input
meshes (the "Chair" example), hampering performance. 
It also generates a large amount of over-tesselated polygons. 
The "Organic" example is based on a CSG operation with six solids, of the form
$(\mathcal{P}_1 \backslash \mathcal{P}_2) \cup (\mathcal{P}_3 \backslash \mathcal{P}_4) \cup (\mathcal{P}_5 \backslash \mathcal{P}_6)$. QuickCSG directly computes the result, while it is computed with intermediate meshes in~\cite{pavic2010hybrid}, hence the additional second for the intermediate mesh computation.

\subsubsection{Feito et al}

The algorithm from~\cite{feito13} implements two-component 
boolean expressions on triangular meshes, which they resolve by exploring an octree 
with parallel threads.
The authors did not provide their implementation, and thus we compare QuickCSG with the times published in the original paper.
Compared to our test machine, they use a higher-end processor (Xeon X5550 2.7~GHz), with more cores and memory (12~GB).
The evaluation of the original paper relies on combining standard meshes (dragon, armadillo) with a translated version of the same mesh, and averaging the timings results over four CSG operations (union, intersection, and the two possible differences).
QuickCSG is 4 to 5 times faster  for the same core count:  (cr = \# cores):

\begin{center}
\resizebox{\columnwidth}{!}{%
\begin{tabular}{|ll|rrr|rr|}
\hline
Example & $\Nf$ & \multicolumn{3}{c}{\cite{feito13}} & \multicolumn{2}{c|}{QuickCSG} \\
	       &  & 1 cr & 4 cr  & 16  cr& 1 cr & 4 cr \\ 
\hline
Armadillo & $2\times150$k & 2.71 & 1.46 & 0.68 & 0.57 & 0.24 \\
Dragon & $2\times871$k  & 12.64 & 6.48 & 2.72 & 2.61 & 1.18  \\
\hline
\end{tabular}%
}
\end{center}

Possible reasons for this performance gap could come from  the intermediate geometry generated  (over-tesselized triangles and vertices that must be merged in a later stage), and  an octree fully stored in memory as it is required for  ray shooting, unlike our approach which doesn't require storing the tree.%

\subsection{Performance \& Comparisons on Huge Datasets}
\label{sec:datasets}

\begin{figure*}
\begin{center}
\resizebox{\linewidth}{!}{
\begin{tabular}{|ccc|}
\hline
\multicolumn{3}{|c|}{Input} \\
T1 & T2 & H \\
$\N=50$, 40k vertices, 40k facets & 
$\N=50$, 3500 vertices, 3500 facets &
$\N=42$, 16k vertices, 33k facets \\
\includegraphics[height=4cm]{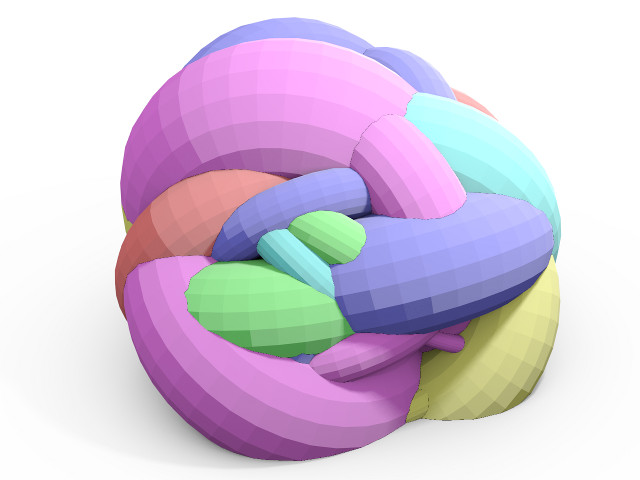} & 
\includegraphics[height=4cm]{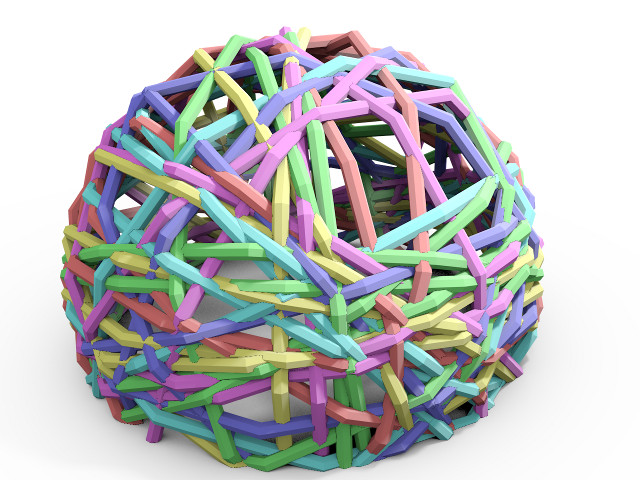} & 
\includegraphics[height=4cm]{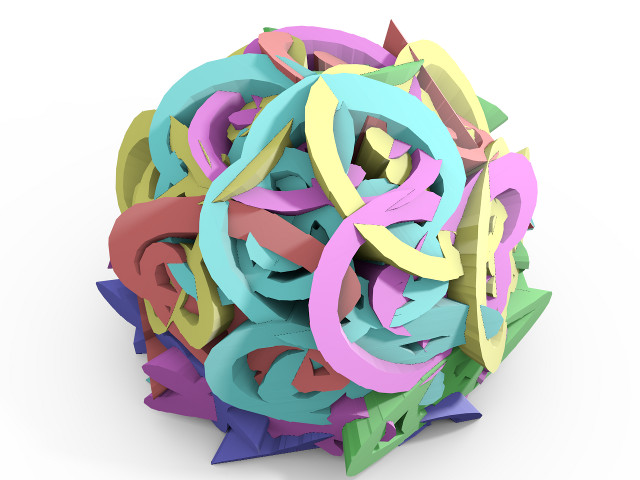} \\
\hline
\hline
\multicolumn{3}{|c|}{Ouput (for vertices, we indicate: \#order-1 + \#order-2 + \#order-3 = total \# vertices)} \\
9207 + 6116 + 372 = 16k vertices, 32k facets & 
699 + 38220 + 7876 = 47k vertices, 94k facets & 
0 + 16508 + 10728 = 27k vertices, 14k facets \\
\includegraphics[height=4cm]{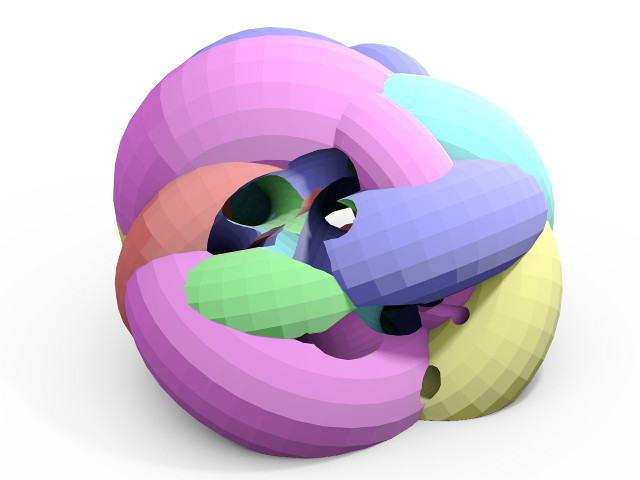} & 
\includegraphics[height=4cm]{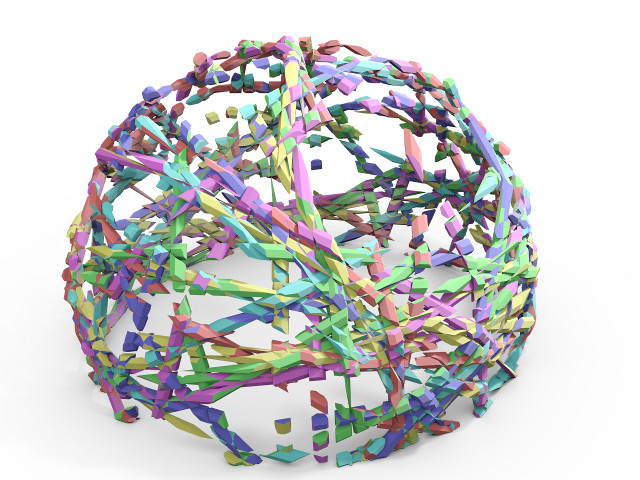} & 
\includegraphics[height=4cm]{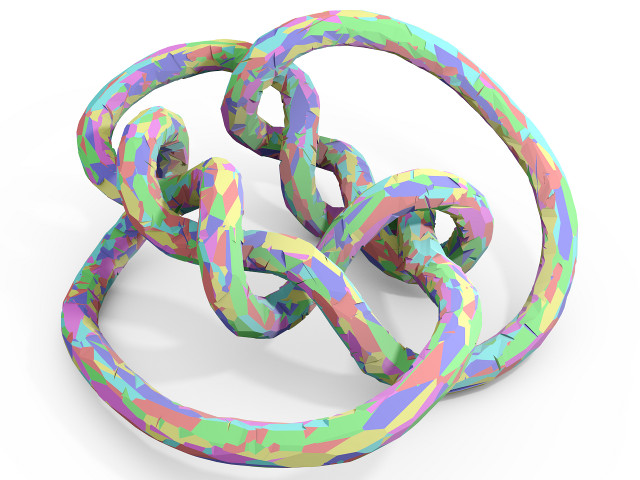} \\
\hline
\hline
\multicolumn{3}{|c|}{QuickCSG runtime (topology + \textsc{CSGVertices} + \textsc{CSGFacets} = total) and memory usage} \\
 7.1   + 69.9  + 10.4 = 87.4 ms &
 0.9   + 78.3  + 72.5 = 151.7 ms &
 5.1   + 437.6 + 26.2 = 468.9 ms \\
61 MB RAM & 
45 MB RAM &
96 MB RAM \\
\hline
\end{tabular}
}
\end{center}
\caption{ The {\bf T1}, {\bf T2} and {\bf H} test cases and QuickCSG performance. RAM is measured as the maximum resident set size reported by the unix \texttt{time} utility. 
}\label{tab:testdatasets}
\end{figure*}

Importantly, all previous tests were performed on the datasets of the original papers, which only process a few, sparse-intersecting inputs, and a small number of boolean operations,~\ie in the ``comfort zone'' of existing algorithms. The algorithm is already shown to significantly outperform them in this favorable situation. To illustrate the even larger gain in the more general situations our algorithm can tackle, we introduce new test cases involving large meshes, very dense intersection areas, with several dozen operations:
\begin{itemize}
\item 
 {\bf T1} is a set of 50 random toruses. We compute the difference between the
  union of the 25 first toruses with the union of the 25 next ones:
  $\Pres = (A_1 \cup \cdots \cup A_{25}) \backslash (A_{26} \cup
  \cdots \cup A_{50})$. This is a typical CSG case, where 
  many facets intersect, but the geometry is regular (small
  compact facets). 

\item 
  {\bf T2} is  a set of 50 concentric narrow random toruses. The toruses follow 
  the great circles on a sphere, so each torus intersects each other
  torus in two locations. We compute the volumes where at least two of
  the toruses are present ($f_\mathrm{min-2}$). In this case, most
  facets intersect another. This generates many disconnected components 
  with many more facets than there are on input. 

\item 
  {\bf H} is  a set of 42 cones with arbitrary bases corresponding to the silhouettes of
  a piece of rope seen from 42 cameras. The silhouettes define cones whose apexes are the 
  optical centre of the cameras, and that pass through the silhouette's shape on the camera's image planes. 
  An approximation of the piece of rope can be reconstructed by intersecting ($f_\mathrm{\cap}$) the cones~\cite{FrancoEPVH}. 
  The facets are very elongated, and the output  mesh has  no order-1  vertices. 
\end{itemize}

Table~\ref{tab:testdatasets} gives some statistics about these three test cases  and QuickCSG's performance.
 The performance timings are broken down in three stages: the ``topology'' concerns a necessary preprocessing pass of our algorithm over the data for facet normals and adjacencies, and the other timings report the execution time of \textsc{KDVertices} and \textsc{CSGFacets}. The
behavior can be different depending on the mesh. For T1 and H, the
slowest stage is \textsc{CSGVertices}, for T2 it is~\textsc{CSGFacets},
because it generates many facets.

\subsection{Comparisons}

Most existing approaches do not provide their implementation or simply fail in this case. For example, 3DSMax ProBoolean produced an incorrect result for T1, after 12~s of computation. Therefore, we compare results of T1, T2 and H, against one of the fastest and most robust method available, the Carve library. The operations on T1 and H were expressed as trees of
binary operations  and the operation on T2 as the union of all intersections of 2 meshes 
to enable Carve to process them.  This last  operation has  $\scriptsize\left(\begin{array}{c}n\\2\end{array}\right)$ components:
\begin{equation}
\begin{array}{ll@{}l@{}l@{}l@{}l}
\multicolumn{4}{l}{f_{\mathrm{min}-2}(a_1,\cdots,a_n) =} \\
\hspace*{3em}&f_\cup(& f_\cap(a_1, a_2), & f_\cap(a_1, a_3), & \cdots, & f_\cap(a_1, a_n), \\
&         &                        &  f_\cap(a_2, a_3), & \cdots, & f_\cap(a_2, a_n), \\
&         &                        & \cdots  & & f_\cap(a_{n-1}, a_n)) \\
\end{array}
\label{eq:min2binary}
\end{equation}

Sequential execution times are found to be:

\begin{center}
\begin{tabular}{|lr|rr|}
\hline
Example & $\Nf$ & Carve CSG & QuickCSG \\
\hline
T1                                 & 40000  & 4.652 &  0.297  \\
T2                                 & 3500 & (94.795) & 0.596 \\
H                                  & 33108  & 26.330 &  1.720\\
\hline
\end{tabular}
\end{center}
\noindent
Times are given in parenthesis when  Carve was only able to provide a partial result. 
For T1 and H, both QuickCSG and Carve provide the expected results,  QuickCSG being about 15 times faster than Carve. 
For T2, despite its  careful handling of degenerate cases, Carve was not able
to compute $f_\mathrm{min-2}$ on more than 38 input meshes. Indeed, computing the unions of the intermediate intersections   generates many degeneracies because Carve relies on several binary operations of meshes that use exactly the same vertices but with different incidence, which fails for non-exact methods such as Carve.

\begin{figure}
\begin{center}
\includegraphics[width=0.7\linewidth]{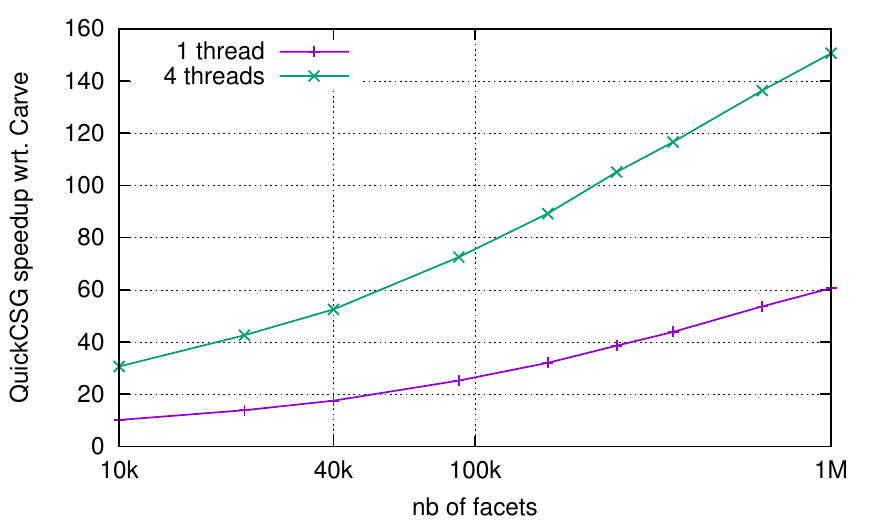}
\end{center}
\caption{\label{fig:t1subdiv}
	Speed comparison between Carve and QuickCSG for meshes of increasing complexity based on T1 (which corresponds to the point at 40k facets). 
}
\end{figure}

Additionally, we evaluated how the speedup evolves as a function of size of input meshes, by generating increasingly subdivided versions of the T1 dataset. We report timings in Figure~\ref{fig:t1subdiv} against runs of Carve on the same datasets. Carve fails over one million input polygons in this example. Speedups for QuickCSG: with 100k polygons, mono-core QuickCSG is $30\times$ faster than Carve, and 4-core executions are $70\times$ faster. With one million input facets, this speedup reaches $60\times$ mono-core and $150\times$ with 4 cores.

\subsection{Probing the Characteristics of QuickCSG}

We validate some  properties of the algorithm  to  support the claims 
of the previous sections: namely, it is efficient to directly compute the result of CSG operations on multiple polyhedra, 
the complexity of the CSG operations is $\mathcal{O}(m\log h)$, and  the parallelization is efficient.

\subsubsection{Comparison with binary CSG operations}
\label{sec:cmpbinary}

We evaluate  the performance improvement  that QuickCSG can provide by directly computing the final output without 
 intermediate polyhedra, compared to the classical approach that  combines polyhedra by pairs. Indeed,
any associative boolean operation $f$ over $n$ input polyhedra can be expressed as a sequence or binary tree
of $n-1$ binary operations:
\begin{equation}
\begin{array}{rcl}
f(a_1,\cdots,a_n) &=& f(a_1, f(a_2, \cdots f(a_{n-1}, a_n ) \cdots ) ) \\
 &=& f( \cdots f( a_1, a_2) \cdots , \cdots  f(a_{n-1}, a_n ) \cdots ) ) \\
\end{array}
\label{eq:binop}
\end{equation}
Such decompositions produce $n-2$ intermediate polyhedra.

Table~\ref{tab:cmphierarchy}  compares the execution for  different combinations. 
The CSG operations on T1 and H can be expressed with binary operations applied sequentially 
or through a binary tree.  Directly evaluating the equivalent  $n$-ary operation to 
compute the   output polyhedron clearly outperforms binary operations that produce many facets of 
intermediate polyhedra that are discarded later on.   Producing  intermediate  polyhedra also increases the
probability of errors caused by degeneracies.

We investigate  alternate trees of operations by varying the arities at different levels.
In Table~\ref{tab:cmphierarchy}, for example ``25,2'' in T1 means that we first compute 
the union of  25 polyhedra  and next combine the resulting two polyhedra to get the final result. %
The results shows that it is generally faster to perform the CSG operation in one pass, except 
for T1, where the ``25,2'' ordering gives the best results. In this case the reason could be  
that operating on fewer meshes at a time means that fewer
vertex/cell indicator bits need to be computed for unrelated components.

\begin{table}
\caption{\label{tab:cmphierarchy}
  QuickCSG execution times when expressing the same CSG operation differently.  Tests performed on the  T1 and H test cases with 4~threads. Execution times  slightly  differ  from Figure~\ref{tab:testdatasets}  because they are obtained via the Python interface of QuickCSG}%
\begin{center}
\scalebox{0.8}{%
\begin{tabular}{|lrr|}
\hline
\multicolumn{3}{|c|}{T1}\\
\hline
ordering & time & errors \\
\hline
single       & 0.125  & 0 \\
binary tree  & 0.363  & 0 \\
sequential   & 0.619  & 0 \\
5,5,2        & 0.180  & 0 \\
25,2         & 0.111  & 0 \\
\hline
\end{tabular}}%
~~
\scalebox{0.8}{%
\begin{tabular}{|lrr|}
\hline
\multicolumn{3}{|c|}{H}\\
\hline
ordering & time & errors \\
\hline
single       & 0.524  & 0 \\
binary tree  & 2.396  & 9147 \\
sequential   & 2.244  & 50258 \\
8,6          & 0.674  & 0 \\
4,11         & 0.872  & 0 \\
\hline
\end{tabular}%
}
\end{center}
\end{table}

\ifRR
\else

\subsubsection{Evaluation of complexity}
\label{sec:expcomplexity}
We experimentaly probe the expected asymptotical complexity of
QuickCSG, $\mathcal{O}(m\log h)$, by timing  a large number of
random operations. 
We randomly sample 1000 groups of 5 meshes from the ``3D Mesh Segmentation Benchmark''
collection\footnote{\url{http://segeval.cs.princeton.edu/}},
consisting of   379 meshes, from  2600 up to  55k facets each. 
 This collection originally comes from the watertight track of the 2007 shape retrieval 
contest~\cite{giorgi2007shape}.   The CSG operation applied for each
group is  randomly chosen between  union, intersection or union of the 3 first
minus union of the 2 last (labeled diff3 in the figure). 

As in most cases involving complex meshes, the dominant step of the
algorithm is the \textsc{KDVertices} stage, which our experimental analysis thus focuses on.
Even though  the total number of split polygons is found to be $s = \O(\Nf)$, as shown in \fig{segeval} (left), in practice complexity plots are significantly clarified by explicitly accounting for the influence of the number $s$ of splits occurring during the entire exploration, which is volatile from dataset to dataset. Plotting 
the execution time versus $(m + s)\log h$ on the right of \fig{segeval}, shows the proportionality relation between both. The fact that the proportionality is verified over all input sizes and heterogeneous boolean operations brings a strong validation to our analysis of the algorithm complexity in Section~\ref{sec:complexity}.

\begin{figure}
\begin{center}\makebox[0pt]{
\includegraphics[height=3.5cm]{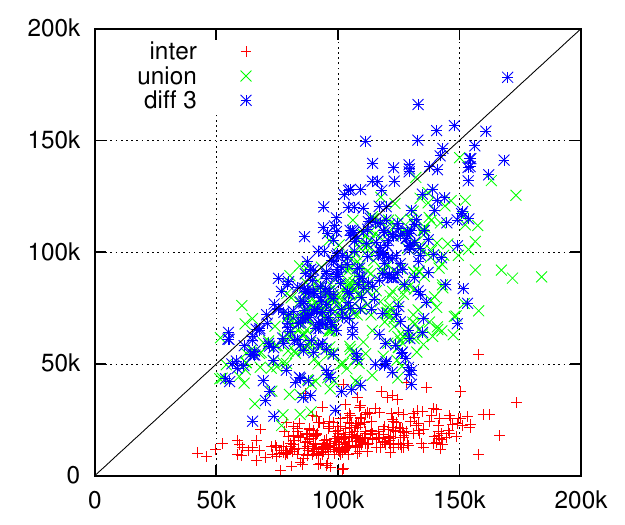}%
\includegraphics[height=3.5cm]{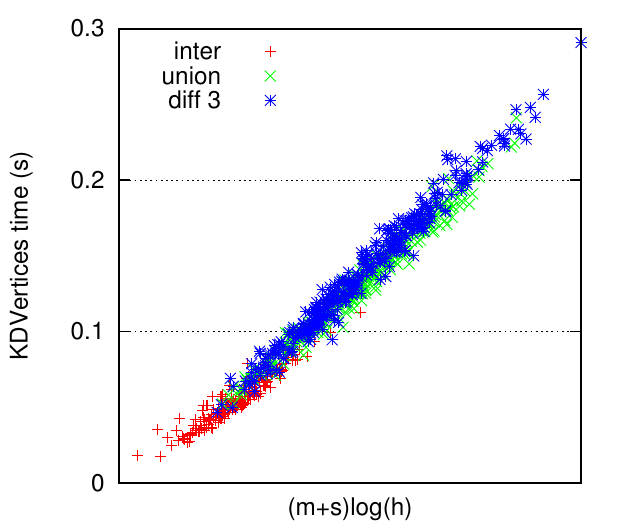}}
\end{center}
\caption{\label{fig:segeval}
  Test on 1000 CSG operations on 5 random meshes. 
  Left: Relationship between the number of split polygons ($s$, y-axis) and the number 
  of input facets ($m$, x-axis). Right: the runtime for the main computation as a function of $(m+s)\log h$.
}
\end{figure}

\fi 

\subsubsection{Parallel Execution}

\begin{figure}
\includegraphics[width=\linewidth]{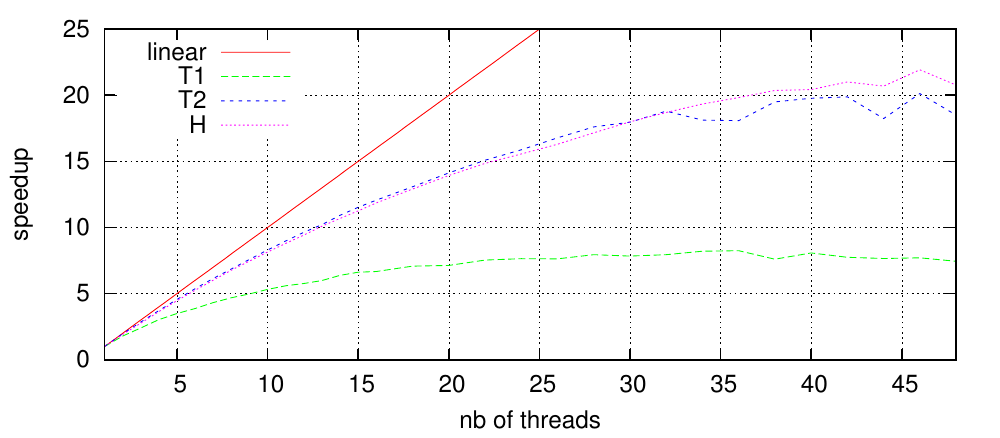}
\caption{\label{fig:threadspeedup} Speedup obtained with more threads,
  with respect to a sequential run. This is measured on a 48-core AMD
  Opteron(tm) Processor 6174. 
}
\end{figure}

We evaluate QuickCSG's parallelization on 
 a machine with  four  12-core AMD processors. 
Figure~\ref{fig:threadspeedup} plots the speedups  for the T1, T2 and H
tests. The Gantt charts in  Figure~\ref{fig:gantt} give a detailed insight
about the parallelization behavior at 48 cores. Despite the irregular nature
of the computations, parallelization is more than   $80\%$  efficient up
to 8 cores for T1 and 20 cores for T2 and H.  At large core count, the performance
is impacted by the exploration of the top KD-tree levels and the data gathering  before 
building  facets.
The parallelization efficiency  depends on the complexity of the
input and output meshes. For instance the T1  output consists of a large majority of first order vertices, with few order-2 and order-3 vertices, so there are too few \textsc{CSGVertices} tasks in the KD-tree for the distribution over so many cores to be efficient, see Figure~\ref{fig:gantt}. The parallel performance improves for large meshes with complex outputs.

\section{Applications}
\secl{applications}

In this section we demonstrate several use cases where the performance of QuickCSG opens new possibilities for usage of boolean combinations of polyhedra. We first examine the computer vision problem of 3D modeling from a set of silhouettes, then solid modeling in the context of 3D printing, and collision detection in interactive systems. We conclude the section on extreme uses of boolean operations, in real-time or over million-polygon inputs.

\subsection{3D Modeling}
\label{sec:3drecons}

\begin{figure}
\includegraphics[width=0.6\linewidth]{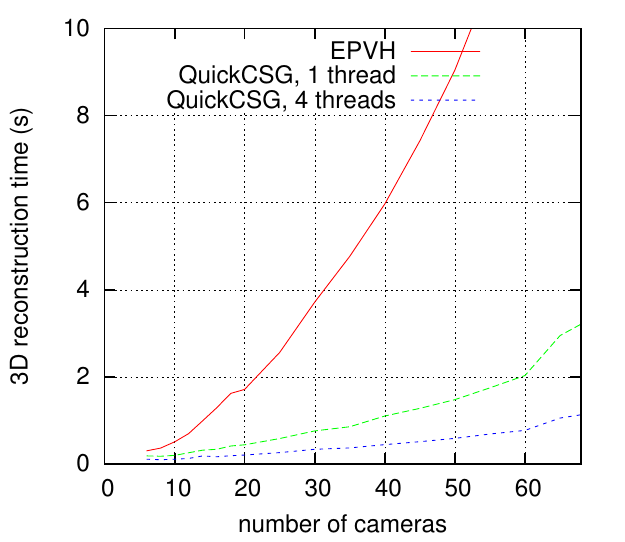}\!\!\!%
\includegraphics[width=0.5\linewidth]{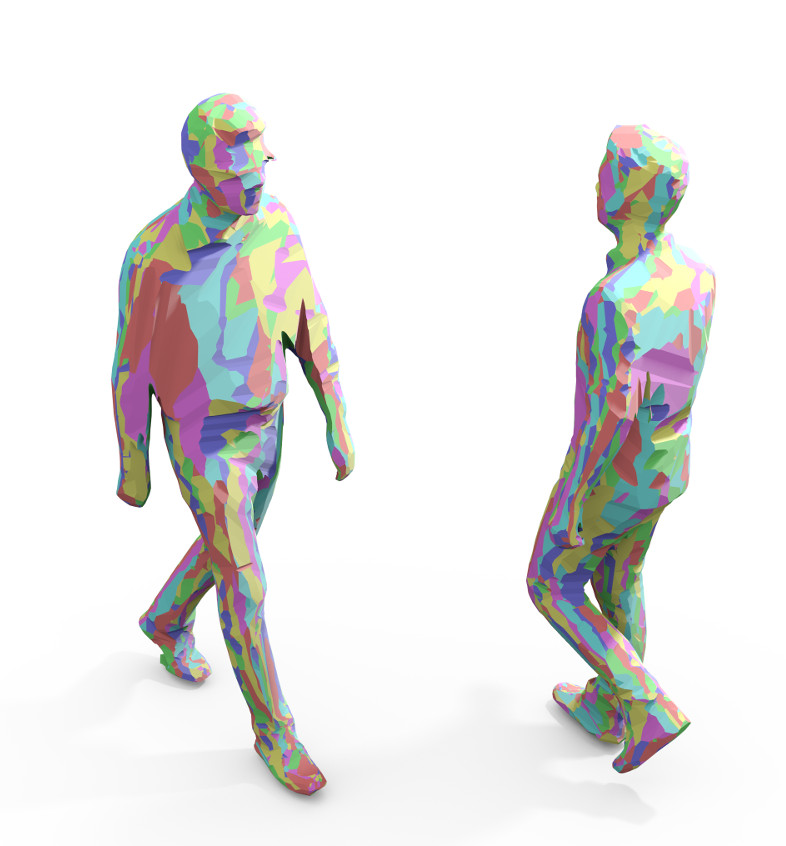}
\caption{\label{fig:cmpepvh}
  (Left) QuickCSG vs. EPVH runtime.  (Right) Resulting visual hull when computed from 68 cameras  (22k facets). 
}
\end{figure}

Given $n$ real photographic frames of an object acquired from different camera viewpoints, it is possible to build 3D reconstructions of the object by extracting its silhouettes in the obtained images, and building the \emph{visual hull} of the object. The visual hull is the maximal 3D volume that projects onto the input silhouettes. Previous works have shown that it can be built by intersecting a set of polyhedral viewing cones~\cite{baumgart74}, \ie cones whose apex is the optical center of the cameras, and whose basis is the silhouette itself. Because the quality of the model increases with the number of input views, new dedicated multi-camera acquisition platforms are being crafted with several dozen cameras, as is the case for example of the Kinovis  68-camera studio\footnote{\url{http://kinovis.inrialpes.fr}}. These platforms provide huge amounts of complex data, challenging even the fastest existing implementations.
In this context, we compare our algorithm with one of the fastest state-of-the-art methods specialized in this task, EPVH~\cite{FrancoEPVH}. We use a 68-camera dataset produced on the Kinovis platform, plotting the execution time of both methods against the number of input cones, see Figure~\ref{fig:cmpepvh}. QuickCSG significantly outperforms the dedicated method whatever the number of input views considered, reaching a ten to twenty-fold speed increase above 30 cameras with four threads.

\subsection{Solid Modeling} \label{sec:cmpopenscad}

\begin{figure}
\resizebox{\columnwidth}{!}{%
\begin{tabular}{c@{}c}
\multicolumn{2}{c}{\includegraphics[width=\linewidth]{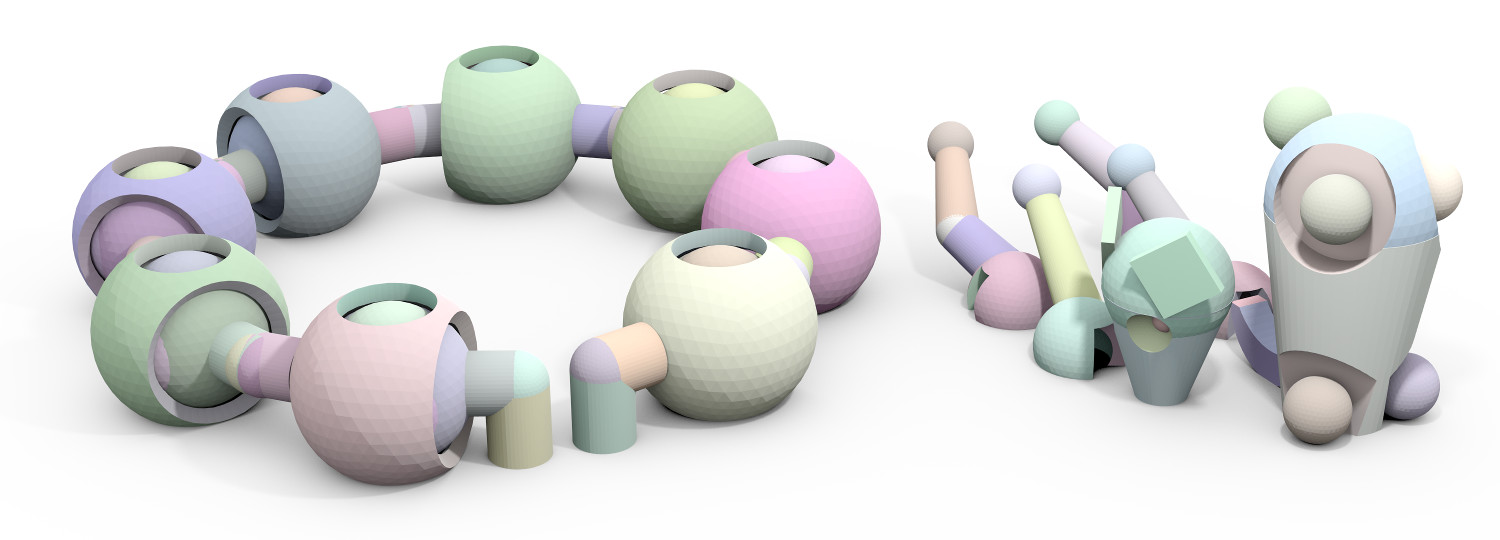}} \\
Balljoint (154 primitive objects) & Doggie (51 primitive objects)\\
\includegraphics[width=0.5\linewidth]{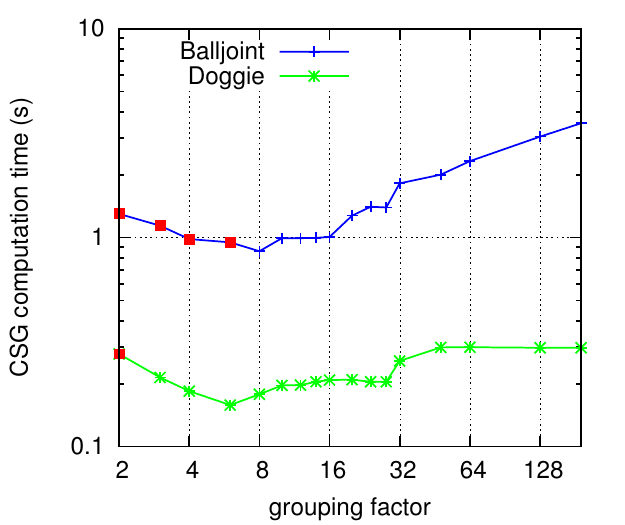} &
\includegraphics[width=0.5\linewidth]{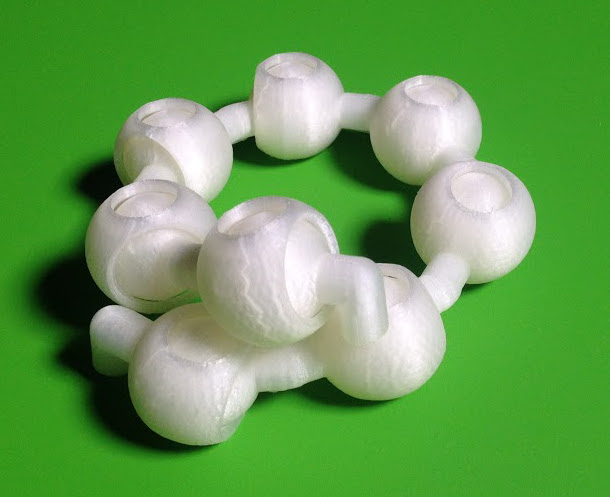} \\
\end{tabular}%
}
\caption{\label{fig:openscad}
  Top: The Balljoint and Doggie models. Down left: QuickCSG runtime (single thread) with different grouping strategies (red squares indicate that there were degeneracy errors). Down right: printed Balljoint. All balljoints are functional.
}
\end{figure}

With the increasing popularity of 3D printing, there is a demand for fast and convenient modeling tools that enable any user to sketch and put together his own 3D objects for printing. %
OpenSCAD\footnote{\url{http://www.openscad.org/documentation.html}} is a popular software and scene description language used in this context. Solids are modeled through a  tree of  binary CSG  operations from primitive shapes (cylinder, sphere, extruded 2D shapes, etc.). OpenSCAD relies on CGAL~\cite{hachenberger07} to  compute  the resulting polygonal mesh.  We compare the performance of QuickCSG versus OpenSCAD when both process the same binary tree of operations. 
Tested on two complex models, Balljoint  and  Doggie from IceSL~\cite{lefebvre.13.aefa},  OpenSCAD's mesh computation  requires 16 minutes and 7 minutes respectively while  QuickCSG's only needs 1.46~s and 0.3~s. We printed the Balljoint mesh generated by QuickCSG using  Makerware on a Makerbot Replicator~2 (Figure~\ref{fig:openscad}). All the balljoints are functional.

It is also possible to compute the whole result at once, by using a boolean function $f$ that evaluates the binary CSG tree. This approach does not perform as well: 3.8~s for Balljoint. 
To evaluate the trade-off between binary and all-at-once computation, 
we traverse the OpenSCAD CSG tree, returning an intermediate sub-tree for each node.
 For a given node, we collect the result sub-trees of its two child nodes. If the total number of meshes in these sub-trees is above some threshold $G$ (the grouping factor), we call QuickCSG to compute the CSG operation and return a 1-node sub-tree with the CSG result. Otherwise, we return a sub-tree built with the two child mesh results and the binary operation. The two baselines, binary evaluation and all-at-once evaluation are obtained  for $G=2$ and $G=\infty$ respectively. 

Figure~\ref{fig:openscad} plots the execution time as a function of the grouping factor $G$.  The optimum occurs at $G=8$, with a  30\% gain  compared to the binary evaluation: 0.96~s for Balljoint and 0.2~s for Doggie (or 0.52~s and 0.12~s respectively with 4~threads). 
The binary tree built by OpenSCAD  already conveys  some spatial  clustering. Two intersecting primitive shapes are very likely close in the binary tree. 
By grouping operations according to their location in the binary tree, we indirectly benefit from a better space partitioning than the one the KD-tree performs 
when considering all operations at once. 

QuickCSG is 500 to 1000 faster than CGAL. Because of the perfect CAD inputs, which exhibit more regularity than acquired datasets, the mesh produced by QuickCSG occasionally suffers from localized degeneracies, even with jittering, unlike CGAL that relies on exact methods. However, when dealing with higher polygon counts, CGAL computation times quickly becomes impractical. The huge performance gap offered by our algorithm leaves room for building an exact version of the method with still significantly faster runtimes.

\subsection{Collision Detection}

\begin{figure}
\begin{center}
\framebox{\includegraphics[width=0.4\linewidth]{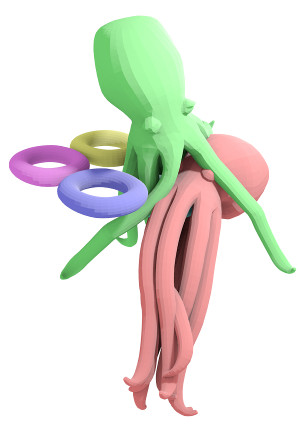}}~~~~~%
\framebox{\includegraphics[width=0.4\linewidth]{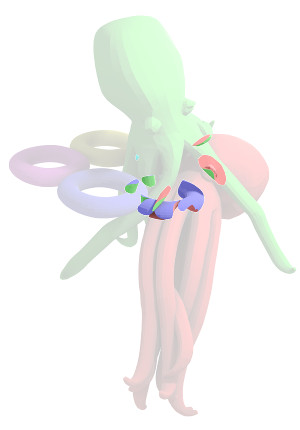}}
\end{center}
\caption{\label{fig:octopus}
	QuickCSG computes min-2 on two flabby octopuses and four rings. (Left) 33k input faces in total. (Right) Interpenetration volumes.
}
\end{figure}

Many interactive systems rely on virtual objects simulations that necessitate inter-object collision detection. A vast array of dedicated methods have been designed just for this problem, often based on hierachical structures that reduce the otherwise quadratic object interpenetration tests~\cite{weller13}. Interestingly however, the existence of generic high-performance CSG tools can provide a new basis to reformulate the problem. Given a set of solids, the set of object interpenetrations subvolumes can be obtained by computing the min-2 operation over all objects in a given scene. Each connected component of the output is the interpenetration volume of at  least two solids (this is approach is as yet incomplete, since it does not handle self-intersections and interpenetrations of more than 3~solids). 
We ran QuickCSG on an example of the SOFA physics engine\footnote{\url{http://www.sofa-framework.org}.}, see~\fig{octopus}. QuickCSG on 4 cores takes 15.5~ms (excluding the 6~ms topology pass, which can be run once at the beginning of the animation), while the state-of-the-art LDI  method from~\cite{allard2010contact} computes collisions and interpenetration volumes  in 5~ms  on a GPU (Quadro 4000). Thus, our generic CPU-only implementation is only a 3-4 factor away of a specialized approximate algorithm running on dedicated hardware. %

\subsection{Extreme CSG}

\label{sec:extremecsg}

\begin{figure}
\begin{center}
\begin{tabular}{cc}
union & min-2 \\
\includegraphics[width=0.5\linewidth]{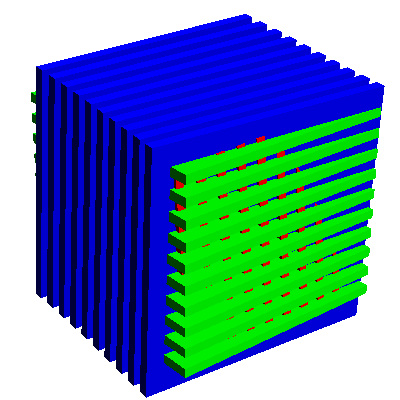} &
\includegraphics[width=0.5\linewidth]{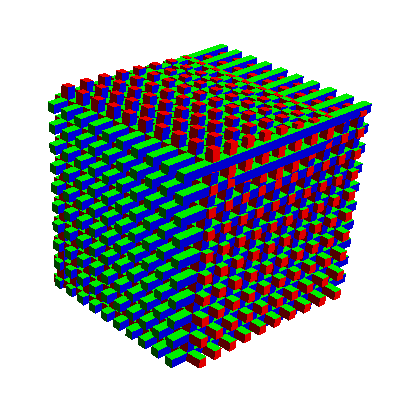}\\
\end{tabular}
\end{center}
\caption{\label{fig:showcase}
	Union and min-2 running at 30fps on  $3\times 10$ undulating boxes, generating 17k and 27k facets respectively. The number of output facets $h$ grows as $h \sim m^3$ with the number of input facets $m$.
}
\end{figure}

The speed of QuickCSG makes it well suited for interactive applications.  We developed a small Python OpenGL application that animates a set of moving or deforming polyhedra, and combines them with a user-specified boolean function. Performing the min-2 or union of three sets of quasi-parallel boxes (see~\fig{showcase}), runs at 30~fps  for $3\times 10$ boxes, while the output  has a large number of triangles ($10\times10\times10$ grid of 3D crosses for min-2).

\begin{figure*}
\begin{center}
\includegraphics[width=17cm]{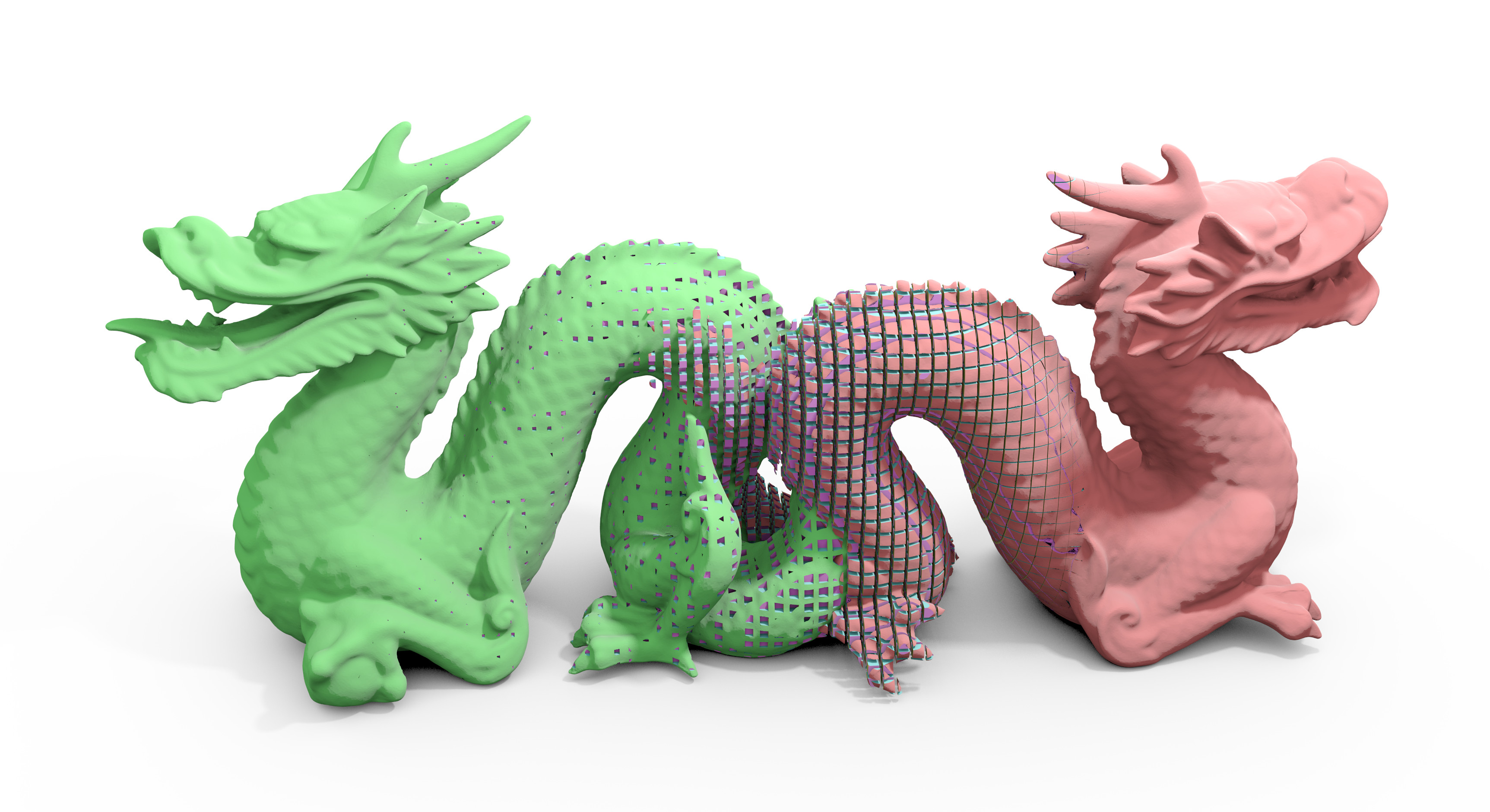}
\includegraphics[width=15cm]{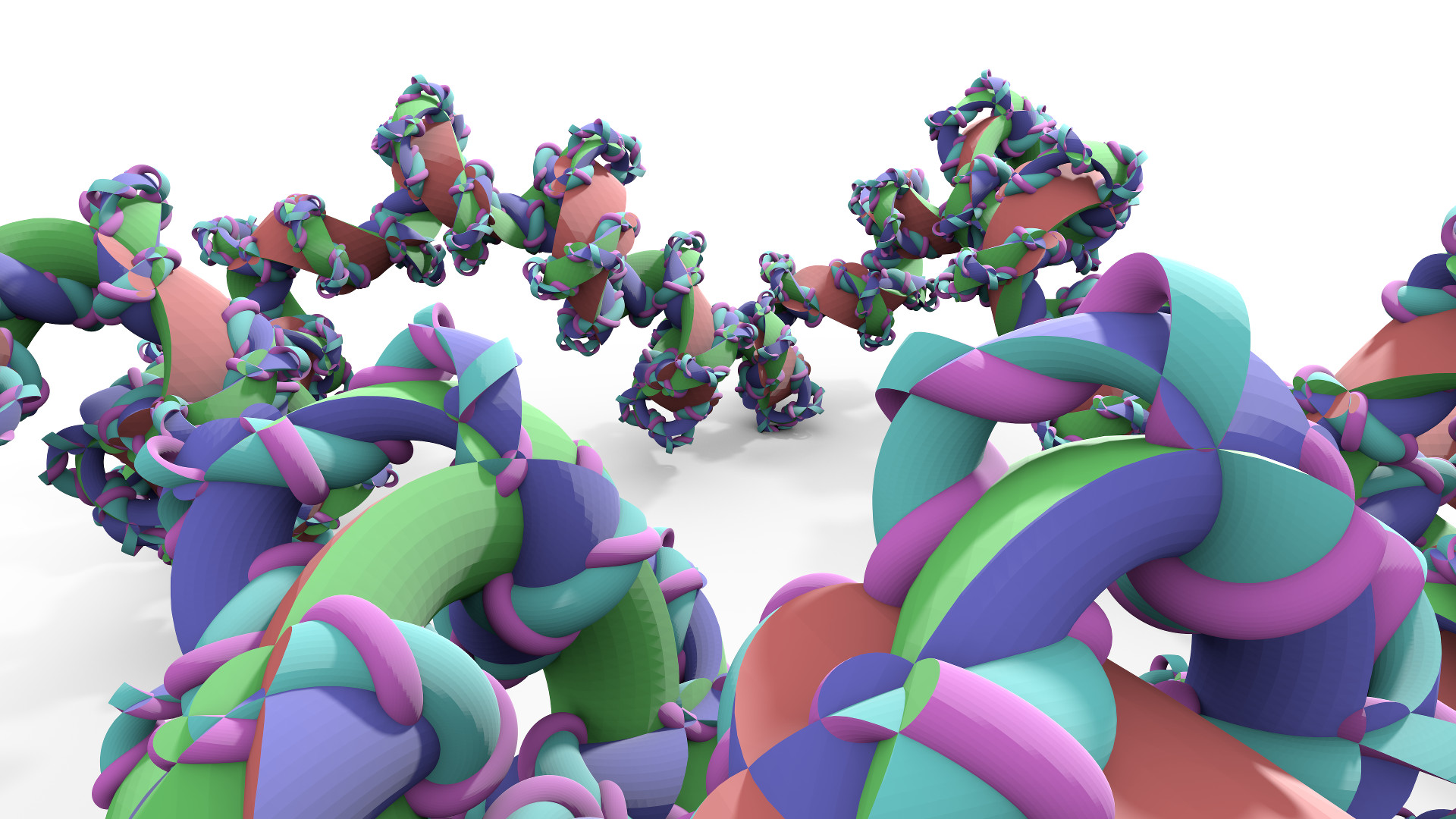}
\end{center}
\caption{\label{fig:results}
   The  Dithering (1.5M triangles) and Serpent (10M triangles) tests.  
}
\end{figure*}

To give an idea of the broader applicability of the method on million-polygon datasets, we test QuickCSG on huge and intrinsically dense and complex examples, see Figure~\ref{fig:results}. 

The  Dithering test  mixes two dragon meshes $\P[1]$ and $\P[2]$ with a 3D dithering pattern. The pattern is defined as the union of three orthogonal combs: $D=\P[3] \cup \P[4] \cup \P[5]$.
Then the dragon meshes are combined using the pattern as a mask: $\Pres = (\P[1] \cap D) \cup (\P[2] \backslash D)$. This is a function that would generate degeneracies by definition, if evaluated as a CSG tree. Indeed, in the intersection volume $\P[1] \cap \P[2]$ it computes $D \cup \neg D$. 
The result is computed in 2.5~s on our test machine (input: 1.74M facets, output: 1.69M). 

The Serpent dataset is another such example, built as a fractal, where a tube, $\P[1]$, is wound around a torus. Then another tube, $\P[2]$, winds around $\P[1]$ and so on until $\P[5]$. We compute the min-2 operation. From 31M input facets, QuickCSG outputs a~10M triangle mesh with a topological genus~\cite{compgraphgeommodel} of  701, i.e., it can be transformed without tearing into a sphere with this many handles. This dataset overwhelms the memory of our standard test machine, so we computed it on a 12-core Mac Pro machine with 64~GB of RAM in only 15 seconds. 

The last example is built from six instances of the Happy Buddha mesh~\cite{happybuddha96}, the largest mesh from the Stanford 
repository. We intersect these with the union of 100,000 random spheres. 
The spheres were labeled with a greedy graph coloring algorithm to group them into 37 disjoint
subsets, so there are a total of 43 input meshes and 24M
triangles. The CSG operation computes the union of the 6 Buddhas and intersects this with the 
union of all spheres: $\Pres=(\mathcal{P}_1 \cup \cdots \cup \mathcal{P}_6) \cap (\mathcal{P}_7 \cup \cdots \cup \mathcal{P}_{43})$. On the Mac Pro this last example runs in  8~s and generates 5M triangles. It is shown on Figure~\ref{fig:6buddha}.

\section{Discussion \& Conclusion}
\secl{conclusion}

We have presented a new output-sensitive approach to boolean solid modeling, which generalizes previous known methods to the N-polyhedron case with arbitrary expressions, by directly computing the result with a simplified, vertex-centric algorithm. Thanks to its straightforward divide-and-conquer and pruning scheme, the algorithm achieves a performance breakthrough on datasets of all sizes. This has been extensively verified experimentally against a vast array of state-of-the-art approaches. The speedup not only materializes on typical cases previous algorithms would be used in, it proves to be groundbreaking on extremely dense and large datasets with up to tens of million polygons. The performance speedup in these cases reaches up to three orders of magnitude with respect to commonly available approaches, when the latter do not fail due to the inability to deal with the large data.

The efficiency and expressiveness of the approach opens new possibilities as it also enables computation of results for arbitrary expressions, some of which simply cannot be processed with existing approaches. For this reason we believe many use cases of boolean modeling for research problems, initially ruled out for feasability and performance reasons, now become accessible. 
We have shown some possible applications of the algorithm in the context of solid modeling for 3D printing, computer vision, interactive systems. We make the program, datasets, experimental protocol, and additional results available to the research community in the supplementary material and on the following page: \url{http://kinovis.inrialpes.fr/static/QuickCSG}.

Many future developments and ramifications of this method are possible. First, the demonstration here was done with polyhedral solids, but the method could be extended to other B-Rep representations, such as parametric surfaces, as the topological analysis shown in the paper is identical with curved facets, edges, and vertices at the intersection of curved edges and facets. Second, although the emphasis here is on performance with fast but non-robust predicates, one can imagine deriving an exact version of the algorithm relying on the Simulation of Simplicity paradigm~\cite{edelsbrunner90}, as the algorithm relies on a small number of well identified core geometric constructs and predicates (rayshooting in \textsc{isFinal}, trihedron orientation test in \textsc{CSGVertices}, intersection routines \textsc{Intersect2Facets}, and \textsc{intersectSegmentFacet}, vertex ordering in \textsc{First}, axis-aligned plane splitting in \textsc{Split}). The inherent property of the algorithm to compute only final geometric primitives and no intermediate results will necessarily benefit this use case, as a substantial fraction of the speed penalty in converting to exact predicates would thus be avoided. Finally, the fact that the algorithm improves the known upper bound in complexity of boolean solid operations raises the broader theoretical question of its optimality for this problem, which we will investigate in future work.

\bibliographystyle{acmtog}
\bibliography{mCSG}

\end{document}